\documentclass[preprint]{aastex}
\usepackage{natbib}
\usepackage{paralist}
\usepackage{adjustbox}
\usepackage{graphicx}

\usepackage[normalem]{ulem}

\shorttitle{Photoevaporation Does Not Create A Pileup}
\shortauthors{Wise \& Dodson-Robinson}

\begin{document}


\title{Photoevaporation Does Not Create a Pileup of Giant Planets at 1 AU}


\author{A. W. Wise\altaffilmark{1,2} and S. E. Dodson-Robinson\altaffilmark{1,3}}
\affil{University of Delaware, Newark, DE 19716}

\altaffiltext{1}{University of Delaware, Department of Physics and Astronomy, 217 Sharp Lab, Newark, DE 19716, USA}
\altaffiltext{2}{Email: aww@udel.edu}
\altaffiltext{3}{Email: sdr@udel.edu}

\clearpage

\begin{abstract}

The semimajor axis distribution of giant exoplanets appears to have a
pileup near 1~AU.
Photoevaporation opens a gap in the inner few AU of gaseous disks before dissipating them. Here we investigate whether photoevaporation can significantly affect the final distribution of giant planets by modifying gas surface density and hence Type II migration rates near the photoevaporation gap. We first use an analytic disk model to demonstrate that
newly-formed giant planets have a long migration epoch before
photoevaporation can significantly alter
their migration rates.
Next we present new 2-D hydrodynamic simulations of planets
migrating in photoevaporating disks, each paired with a control
simulation of migration in an otherwise identical disk without
photoevaporation. We show that in disks with surface densities near the
minimum threshold for forming giant planets, photoevaporation alters the
final semimajor axis of a migrating gas giant by at most 5\% over the
course of 0.1~Myr of migration. Once the disk mass is low enough for
photoevaporation to carve a sharp gap, migration has almost completely
stalled due to the low surface density of gas at the Lindblad
resonances.
We find that photoevaporation modifies migration rates so little that it is unlikely to leave a significant signature on the distribution of giant exoplanets.

\end{abstract}





\clearpage

\section{Introduction}


Statistical analyses of the exoplanet mass/semimajor axis distribution
suggest that disk-driven migration plays a critical role in forcing
giant planets ($\ga 0.5 M_{Jup}$) into short-period orbits
\citep{a07,kn12,rc14,s14}. Given that the timescale for Type II
migration---in which a planet opens a tidal gap in the disk
\citep{l96,nelson00,kn12}---is much shorter than observed protostellar disk
lifetimes \citep[$\sim$3~Myr; e.g.][]{h01}, torques from the disk should
have ample time to modify the orbits of gap-opening planets \citep[for
gap-opening criteria, see][]{c06} before disk dissipation by
photoevaporation \citep{h93,h94,c01,f04,a06b,o11,o12,g16}. Indeed, planet
semimajor axis histograms have sometimes been interpreted as showing a
``pileup'' of giant planets with semimajor axes of $\sim$1~AU, or
similarly a planet ``desert'' inside 1~AU \citep{us07,w09,hp12,bn13}.
In this paper we test the hypothesis that photoevaporation
may create a pileup of giant exoplanets near 1 AU by modifying
the disk's surface density, and hence the migration rates of giant plants in the inner disk \citep{m03,ap12,er15}.

Before gas giants can begin Type II migration, they may form with the
help of protoplanetary disk structures called ``planet traps."  First,
planetesimals may grow most easily in local pressure maxima that trap
centimeter to meter-size pebbles, such as near the water ice line
\citep{bs95,l99,b00,cc06,kl07,j09,r13}. With the help of mutual
gravitational attraction, the planetesimals collide to form planetary
embryos \citep[e.g.][]{g78}, which quickly become vulnerable to Type I
migration \citep{gt80,w97}. However, localized disk structures may develop that
balance the migration torques, forming traps that allow embryos to grow
into giant planet cores instead of falling into the star
\citep{m08,s11,hp11}.
Planet cores that rapidly grow to $\ga 10~{\rm M}_\earth$
may begin runaway gas accretion, forming giant planets and
transitioning out of fast Type I migration by carving a tidal gap in the
disk \citep{t07,crida17}. These giant planets' final
orbits will depend on their locations of formation relative to
planet traps, which may leave a signature on the distribution of giant exoplanet locations.



Photoevaporation, in which high-energy radiation from the central star
(or other nearby stars, though we don't consider this case here) drives a disk wind, generates gaps in the inner
few AU of gas disks before dissipating them completely
\citep{c01,a06b,o11}. Planets that migrate into the
ever-widening gap will stop migrating as the disk disperses around
them, possibly creating a pileup in the semimajor axis
distribution of giant planets near the gap-opening radius of
$\sim 1$~AU
\citep{m03,j04, aa09, ap12, er15}.
The disk dissipation induced by
photoevaporation therefore mimics a planet trap. Indeed, Monte Carlo population synthesis models
of giant planets migrating in photoevaporating disks have been shown to
roughly reproduce the observed distribution of giant planets
\citep[e.g.][]{aa09}. \cite{ap12} (hereafter AP12) use an extreme ultraviolet (EUV)-dominated
photoevaporation model to synthesize a population of giant planets with
a desert at $\sim$1-2~AU and pileups on either side (though they
acknowledge the desert location depends sensitively on an uncertain
planetary accretion model), while \cite{er15} (hereafter ER15)  predict a giant planet pileup
between 1 and 2~AU as a result of disk dispersal
triggered by X-ray-dominated photoevaporation. However,
neither AP12 nor ER15 present control simulations where they synthesize
a semimajor axis distribution from migrating planets in
non-photoevaporating disks, so the effects of photoevaporation cannot
easily be disentangled from other parameter choices. While
removing photoevaporation from the AP12 and ER15 simulated disks would 
prevent the disks from ever being completely dispersed, unless
by some other mechanism such as magnetocentrifugal winds
\citep[e.g.][]{gressel15}, the planets' migration rates would
asymptotically approach zero due to the exponentially decreasing
surface density, so non-photoevaporating control simulations
could be constructed that would reveal how gradual gas depletion
might affect the exoplanet semimajor axis distribution.

As our goal is
to isolate the effects of photoevaporation on planet migration, we
directly compare migration tracks of planets in photoevaporating disks
with those of identical planets in otherwise identical, but
non-photoevaporating disks.

This paper is organized as follows.
In \S~2, we briefly describe the models of
disk photoevaporation that we test in our simulations. In \S~3, we
present an analytic comparison of the timescales for viscous disk
evolution, planet migration, and photoevaporative clearing to
demonstrate that giant planets have ample time to migrate before
photoevaporation begins to sculpt the disk. In \S~4, we describe the
setup of our FARGO numerical simulations of planet migration in
photoevaporating disks, as well as the control set of simulations
without photoevaporation. We refer the reader to Appendix~\ref{fargomod}
for a summary of our modifications to the original FARGO 2-D code, and
Appendix~\ref{AppendixParams} for a table of our simulation parameters.
In \S~5, we summarize the results of our simulations and discuss the
extent to which photoevaporation affects planet migration. Finally, in
\S~6, we present our conclusions and ideas for future work.

\begin{table}[t]
\centering
\caption{Symbol Definitions.\label{tbl-1}}
\begin{tabular}{ll}
\tableline\tableline
Symbol & Definition\\
\tableline
$\gamma$ &Adiabatic index of the gas disk\\
$t$ &Age of disk\\
$L_\star$ &Bolometric luminosity of central star\\
$L_\sun$ &Bolometric luminosity of the sun\\
$\dot{M}_{\rm pe}$ &Disk mass-loss rate due to photoevaporation\\
$\dot{M}_a$ &Disk mass-loss rate due to stellar accretion\\
$c_s (r)$ &Disk sound speed\\
$\Sigma (r)$ &Disk surface density\\
$T(r)$ &Disk temperature\\
$\nu (r)$ &Disk viscosity\\
$r$ &Distance from central star\\
$r_{\mathrm{in/out}}$ &Inner or outer radius of disk\\
$\Omega (r)$ &Keplerian orbital angular speed\\
$M_\star$ &Mass of central star\\
$M_d$ &Mass of disk\\
$\mu$ &Mass of hydrogen molecule\\
$t_{\rm pe}$ &Photoevaporation timescale\\
$r_{\rm pe}$ &Radius of minimum photoevaporation timescale\\
$\dot{\Sigma}_{\rm pe}(r)$ &Rate of change of disk surface density due to photoevaporation\\
$a_{\rm start}$ &Starting semimajor axis for migration simulations\\
$\Sigma_{crit} (r)$ &Surface density profile when photoevaporation begins to open a gap\\
$\alpha$ &Viscosity parameter from \cite{ss73}\\
$t_\nu$ &Viscous timescale\\
$L_X$ &X-ray luminosity of the central star\\
\tableline
& \\
\end{tabular}
\end{table}

\section{Photoevaporation Models} \label{photoevap}

In this section we briefly describe the prescriptions for gas removal by
photoevaporation that we use as the basis for our analytic
calculations and simulations. For a discussion of how different types of
ionizing radiation drive disk clearing, see \cite{o10} or \cite{a14}.
For a more detailed discussion of the physics of photoevaporation, we
recommend \S~5 of \cite{a11}.

Photoevaporation is driven by energetic radiation ($h \nu >$ 6 eV)
heating the upper layers of a disk atmosphere so that the sound speed
exceeds the escape speed. A hydrodynamic flow is then launched near the
gravitational radius, $r_g = G M_\star / {c_s}^2$ \citep{h94}. The
hydrodynamic flow is often characterized by the wind-driven mass-loss
rate per unit surface area, $\dot{\Sigma}_{\rm pe}$. When the local viscous
accretion rate falls to ${\sim}\dot{\Sigma}_{\rm pe}$, a gap begins to open in
the disk. Eventually the gap chokes off the accretion flow that supplies
gas to the inner disk (inside the gap), and the inner disk drains on the
viscous timescale. The star then irradiates the interior wall of the
outer disk (outside the original gap) directly, triggering dissipation
on a timescale of ${\la}10^5$~years
(``UV-switch" in the language of \citealp{c01}). This process is
accelerated when a giant planet's tidal gap and the would-be
photoevaporated gap overlap, as the tidal gap and photoevaporation
can both hinder gas accretion to the inner disk \citep{aa09,r13x}.

In our simulations we consider three photoevaporation models for which
$\dot{\Sigma}_{\rm pe} (r)$ or an equivalent expression has been published. The
expressions for $\dot{\Sigma}_{\rm pe}$ used here only apply to disks that have
not drained interior to the planet's orbit (or any photoevaporated gap), i.e., they have not yet
reached the UV-switch (or the equivalent rapid dissipation phase in the
X-ray photoevaporation model). In Appendix~\ref{fargomod}, we describe
how we incorporated photoevaporation into the FARGO planet migration
simulations.

\begin{enumerate}

\item{{\bf EUV}: Hydrodynamic simulations with extreme ultraviolet (EUV)
photoevaporation \citep{l03,f04} show that that flows are actually
launched from roughly (1/5) $r_g$ and the mass-loss rate profile tapers
off quickly at larger radii, resulting in roughly 1/3 of the total
mass-loss rate of the analytic prediction of \cite{h94}.} The EUV
photoevaporation model of \cite{f04} is taken from a numerical fitting
function provided in the appendix of \cite{aa07}.

\item{{\bf X-ray}: \cite{o10,o11,o12} use a similar hydrodynamic
simulation, but consider X-ray, FUV, and EUV fluxes. They find X-ray
photoevaporation to be the dominant driver of disk mass loss, so we
include their X-ray photoevaporation model using the fitting function in
the appendix of \cite{o12}.}

\item{{\bf FUV}: \cite{gh09} took a different approach,
self-consistently modeling the chemical structure of a disk irradiated
by FUV, EUV, and X-ray fluxes, then using the temperature profile found
to estimate photoevaporative mass-loss rates. They found FUV radiation
to be the dominant driver of mass loss. To get their predicted
photoevaporation rates $\dot{\Sigma}_{\rm pe}(t)$, we digitized the solid line
in Figure 2 of \cite{gh09} using an online
app\footnote{http://arohatgi.info/WebPlotDigitizer\label{digitize}}.}

\end{enumerate}

In the next section we present an analytic disk model that suggests
planets will have ample time to migrate before photoevaporation can
significantly affect their migration tracks.

\section{Timescales for Photoevaporation and Giant Planet Migration:
Evidence for a Long Migration Epoch} \label{timescales}


The essential reason photoevaporation cannot significantly affect giant
planet migration is that migration and photoevaporative gap opening
operate at different epochs of disk evolution. (\citealp{h94,c01,a06b}; see review by \citealp{a11}). Since the torque on a
planet from any disk annulus
is proportional to the surface density in that
annulus, migration rates slow as the disk gas accretes onto the star.
Yet photoevaporation only dominates over accretion as the mass-transport
mechanism when the disk surface density has been heavily depleted. We find that
while typical Type II migration timescales are approximately $2 \times 10^5$~years
\citep[e.g.][]{l96,w97,nelson00}, it takes approximately $2 \times 10^6$
years for a disk to deplete from planet-forming densities to low enough
densities for photoevaporation to dominate mass transport, leaving ample
time for newly-formed Jupiters to migrate
unaffected by photoevaporation.


To estimate how much time giant planets have available for migration
before photoevaporation opens a
gap, we construct an analytic disk model that evolves due to viscous forces
and photoevaporation. Our disk initially has roughly the minimum gas
surface density required to form a Jupiter-mass planet ($\sim$500~g~cm$^{-2}$ at 5~AU, see \citealp{l09} and references therein). We let
the disk viscosity $\nu(r)$ follow the $\alpha$-prescription, $\nu(r) =
\alpha {c_s}^2 \Omega^{-1}$ \citep{ss73}. Here $\alpha$ is the viscous
efficiency, $c_s$ is the sound speed, and $\Omega$ is the Keplerian
angular speed for an orbit at radius $r$ around a central star mass of
$M_\star$: $\Omega = \sqrt{G M_\star/r^3}$ (all variables used in this
section are defined in Table \ref{tbl-1}). Our disk is dynamically thin
($H \ll r$, where $H$ is the pressure scale height) but optically thick,
so the sound speed is the adiabatic sound speed, $c_s = \sqrt{\gamma k T
/ \mu}$ (where $\gamma$ is the adiabatic index, $k$ is Boltzmann's
constant, $\mu$ is molecular mass, and $T$ is the local temperature). For
simplicity, we assume the disk gas consists of hydrogen molecules only
so $\gamma = 7/5$ and $\mu$ is the mass of a hydrogen molecule. At each
radius $r$, the disk has a blackbody temperature $T$ in equilibrium with
the stellar radiation field (with bolometric luminosity $L_\star$):
$T(r) = ( L_\star /  (16 \pi r^2 \sigma_{sb} ))^{1/4}$. Combining our
expressions for $T$, $c_s$, and $\Omega$, we can write the
viscosity in terms of parameters $L_\star$, $\alpha$, $M_\star$, and
fundamental physical constants:
\begin{equation}
\nu (r) = \left( \frac{L_\star}{\pi \sigma_{sb}} \right)^{1/4} \frac{\frac{7}{5} \alpha k}{2 \mu (G M_\star)^{1/2}} \ r \equiv \nu_0 \left( \frac{r}{1 \rm{AU}} \right).
\label{eq:visc}
\end{equation}
This simple, physical model of $\alpha$-viscosity recovers the proportionality $\nu(r) \propto r$ as
suggested by \cite{h98}.

Next we want to find the surface
density profile $\Sigma (r,t)$ of a disk undergoing both viscous
accretion and photoevaporation. Unlike \cite{r04}, who find the surface density
evolution in the general case of any disk surface density profile
and any photoevaporation mass loss profile, we
assume a constant value for steady-state accretion onto the star,
$\dot{M}_a (t) = 3 \pi \nu (r) \Sigma (r,t)$ (\citealp{p81}; note this assumption makes
our surface density profile differ from the similarity solution of \citealp{lp74}, but
more closely resemble simulations with a fixed disk outer radius).
We also assume $\dot{M}_{\rm pe} << \dot{M}_a$, where $\dot{M}_{\rm pe}$
is the photoevaporation mass-loss rate (time-independent as long as the
high-energy radiation field is constant and the UV-switch or X-ray
equivalent has not been triggered). Simulations of viscous, photoevaporating
disks tend to exhibit these properties for the majority of the disk lifetime
\citep[e.g.][]{c01,a06b,o11}. These simplifying assumptions give our analytic
model a surface density profile proportional to $1/r$:
$\Sigma(r,t) = \Sigma_{1\mathrm{AU}}(t) ( 1 \mathrm{AU} / r ) $
where $\Sigma_{1\mathrm{AU}}(t)$ is the surface density at 1~AU
from the central star. Under these assumptions, we can write the
total mass-loss rate $\dot{M}_{\rm pe} + \dot{M}_a(t)$
of the disk in terms of the surface density draining rate $\dot{\Sigma}_{\rm 1AU}(t)$:
\begin{equation}
\dot{M}_{\mathrm{pe}} + \dot{M}_a(t) = 2 \pi
\dot{\Sigma}_{1\mathrm{AU}}(t) (1 \mathrm{AU}) (r_{\mathrm{out}}
- r_{\mathrm{in}}),
\label{eq:sigdotmdot}
\end{equation}
where $r_{\mathrm{in}}$ and $r_{\mathrm{out}}$ are the inner and outer
disk radii.
Assuming $r_{\mathrm{in}} \ll r_{\mathrm{out}}$, substituting
the equation for steady-state accretion on the left-hand side of Equation
\ref{eq:sigdotmdot}, and solving the resulting differential equation for
$\Sigma_{1\mathrm{AU}}(t)$ fully specifies the surface density profile
$\Sigma(r,t)$:
\begin{equation}
\Sigma_{1\mathrm{AU}}(t) = -\frac{\dot{M}_{\mathrm{pe}}}{3 \pi \nu_0} + \left(\Sigma_{1\mathrm{AU}}(t=0) + \frac{\dot{M}_{\mathrm{pe}}}{3 \pi \nu_0} \right)e^{-3 \nu_0 t / 2 (1 \mathrm{AU})  r_{\mathrm{out}}}.
\label{eq:sdprofile}
\end{equation}

Now we are able to calculate the critical time, $t_\mathrm{crit}(r)$, at
which photoevaporation begins to contribute significant mass transport
to a given disk annulus,
causing surface density depletion
that cannot be re-filled
by viscous accretion. To do this we set the viscous timescale,
$t_\nu(r) = r^2/\nu(r)$ \citep{p81}, equal to the photoevaporation
timescale,
$t_{\rm pe}(r,t) = \Sigma(r,t)/\dot{\Sigma}_{\rm pe}(r)$,
where $\dot{\Sigma}_{\rm pe}(r)$ is the photoevaporation-induced rate
of change in surface density as a function of disk radius.
Using our forms for viscosity (Equation~\ref{eq:visc}) and surface density,
substituting Equation~\ref{eq:sdprofile} into $t_\nu(r) = t_{\rm pe}(r,t)$
gives us the critical time
\begin{equation}
t_\mathrm{crit}(r) = \frac{-2 (1 \mathrm{AU}) r_\mathrm{out}}{3 \nu_0}\
ln \left( \frac{3 \pi r^2\ \dot{\Sigma}_{\rm pe} (r) +
\dot{M}_{\mathrm{pe}} }{3 \pi \nu_0 \Sigma_{1\mathrm{AU}}(t=0) +
\dot{M}_{\mathrm{pe}} } \right).
\label{eq:tcrit}
\end{equation}
First we analyze this expression by finding the {\it earliest} time that
photoevaporation begins to
dominate mass transport at {\it any} disk radius, or the absolute
minimum of $t_\mathrm{crit}$. The extremal values
of $t_\mathrm{crit}$ occur at radii satisfying:
\begin{equation}
\frac{-2}{r} = \frac{d}{dr} ln( \dot{\Sigma}_{\rm pe}(r)).
\label{eq:extremal}
\end{equation}
For the X-ray photoevaporation model in \S~\ref{photoevap}, we
verified graphically that
Equation \ref{eq:extremal} has only one solution, so we can solve it
with a simple root-finding algorithm. Once time advances to this absolute
minimum value of $t_{\rm crit}(r)$, the condition $\dot{M_{\rm pe}}
\ll \dot{M_a}$ is no longer valid everywhere in the disk and our assumption
of steady-state accretion breaks down, making 
Equations \ref{eq:sigdotmdot} and \ref{eq:sdprofile} no longer
self-consistent.
However, it is instructive to note that for all photoevaporation models
considered (see \S~\ref{photoevap}), the mass-transport timescales
$t_{\rm pe} = t_{\nu}$ found at ($r(t_\mathrm{crit}),t_\mathrm{crit}$)
are much {\it longer} than the rate of change of $r(t_\mathrm{crit})$
(given by the inverse function of Equation \ref{eq:tcrit}) 
at the absolute minimum value of $t_{\rm crit}$.
In other words, $r(t_\mathrm{crit})$ is moving inward {\it faster}
than photoevaporation can drain the disk at $r(t_{\rm crit})$, so a
photoevaporated gap does not form until the inward
propagation of $r(t_{\rm crit})$ slows relative to the draining timescale
$t_{\rm pe}$. For all of the photoevaporating disk models we consider
in the paper, we do not see the gap-opening criterion,
\begin{equation}
\frac{r(t_\mathrm{crit})}{dr(t_\mathrm{crit})/dt} > t_{\rm pe} = t_{\nu},
\label{eq:gapcriteria}
\end{equation}
satisfied until $r(t_{\rm crit})$ moves into the inner few AU of the disk.
The gap center then continues to move at the rate
$dr(t_\mathrm{crit})/dt$ as the gap opens.

Using Equation \ref{eq:gapcriteria}, we now present an analytic estimate
of the time available for planets to migrate before feeling the effects
of photoevaporation. We consider our analytic disk model with parameters
from the well-characterized disk surrounding TW
Hydrae, an old \citep[$\sim$3-10~Myr; see][]{hhp98,vs11} pre
main-sequence star that still has a disk.
This is one of the few disks with a detected photoevaporative
wind emerging from it \citep{ps09,p11}.
For our model of TW Hydrae, we set the current age, mass
and luminosity to be $t_\mathrm{now}=5$~Myr, $M_\star=0.8
\mathrm{M_\sun}$ and $L_\star = 0.25 \mathrm{L_\sun}$ \citep[luminosity
estimated from the stellar evolutionary tracks of][]{dm94}. \cite{b13n}
measure the mass and outer radius of the gas disk around TW Hydrae to be
$M_d = 0.056 \mathrm{M_\sun}$ and $r_\mathrm{out}\approx$~80~AU
respectively. These disk parameters give a surface density of $\sim$850~g~cm$^{-2}$ at 5~AU in our model disk,
which is roughly twice the gas
density required to form Jupiter \citep{j07,l09}. For the mass loss due to
photoevaporation, we use the rates predicted by \cite{o12}, i.e.
$\dot{M}_{\mathrm{pe}} = 6.25 \times 10^{-9}~(M_\star /
\mathrm{M_\sun})^{-0.068}~[L_\mathrm{X}/({10^{30}~\mathrm{erg~s^{-1}}})]^{1.14}~\mathrm{M_\sun
yr^{-1}}$. \cite{rs06} find the X-ray luminosity of TW Hydrae to be
$L_\mathrm{X}=2.0 \times 10^{30}$~erg~s$^{-1}$.  We set $\alpha$=0.001,
on the low end of the expected range from observations (0.01-0.001), to
minimize the time until the photoevaporated gap opens and construct the most optimistic scenario for 
photoevaporation to carve out surface density gradients that alter the planet's migration.
With these model parameters we can write numerical forms for the viscosity
law and surface density profile of TW Hydrae:
\begin{equation}
\nu (r) = 9.0\times 10^{-6} \left( \frac{r}{\mathrm{AU}} \right) \mathrm{AU^2\ yr^{-1}}
\label{eq:TWHvisc}
\end{equation}
\begin{equation}
\Sigma (r,t) = -1.65 \times 10^{-4} + (6.42 \times 10^{-4})\ {\rm exp}\left( -1.7 \times 10^{-7} \frac{t}{\rm yr} \right) \left( \frac{\mathrm{AU}}{r} \right) \mathrm{M_\sun\ AU^{-2}}
\label{eq:TWHsigma}
\end{equation}

\begin{figure}
\plotone{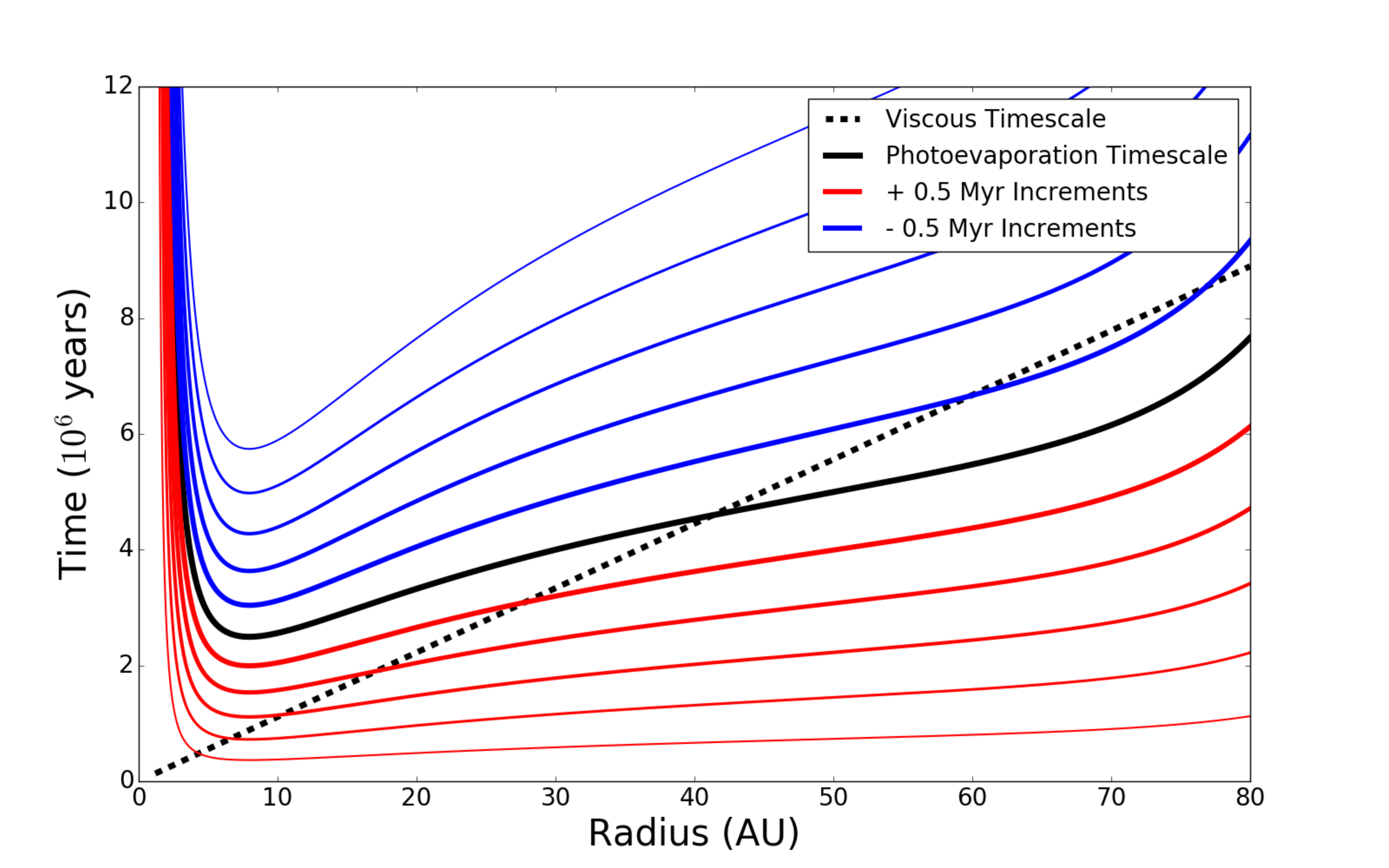}
\caption{
Viscous and photoevaporation timescales as a function of disk radius found by applying our analytic disk model to the disk around TW Hydrae \citep{b13n}. The black dashed line shows the viscous timescale $t_\nu = r^2 / \nu$, and the black solid line shows the present photoevaporation timescale $t_{\rm{pe}}$ (equation 7), and past and future $5 \times 10^5$~year increments are shown in blue and red. Following the intersection of the viscous and photoevaporation timescales shows when and where photoevaporation begins to affect the disk. For the \cite{o12} photoevaporative mass loss profile and the initial disk surface density given by Equation 10, photoevaporation begins to shape the surface density distribution in the inner disk ($r \la 5$~AU) only after about 2.3~Myr of evolution, meaning Jupiter or Saturn analogs could migrate unimpeded by photoevaporation for several migration timescales.
\label{fig1}}
\end{figure}

In Figure~\ref{fig1}, we illustrate the gap-opening process for the
disk around TW Hydrae by plotting the viscous timescale $t_\nu$
and photoevaporation timescale $t_{\rm pe}$ against radius for the
present time, and up to 2.5~Myr in the past and in the future according to
our analytic disk model. Under this model, the gap-moving timescale
$r(t_\mathrm{crit}) / (dr(t_\mathrm{crit})/dt)$ is approximately $10^6$
years when the photoevaporation timescale first drops below the
viscous timescale, which happened roughly 0.6 Myr ago, near 70~AU.
In contrast, the surface density evolution timescale is $t_{\rm pe} =
t_{\nu} = 7 \times 10^6$ years at this time. The gap does not open
until $r(t_{\rm crit}) \la 5$~AU, which agrees well with disk
simulations by \cite{o11} using the same photoevaporation model, as well
as other studies showing a photoevaporation gap-opening radius of
$\sim$1-3 AU \citep{c01,a06b}.


It should be noted that under this model, TW Hydrae's present
photoevaporation timescale $t_{\rm pe} = \Sigma(r,t) / \dot{\Sigma}_{\rm pe}(r)$ is less
than its accretion timescale in the outer disk ($\ga$50~AU), meaning
that the surface density profile in the outer disk may be significantly
modified by photoevaporation. However, we expect that giant planets
formed by core accretion will primarily grow and migrate in the
inner disk. Therefore, to fully specify the time available for a newly-formed
Jupiter to migrate
unaffected by photoevaporation, we calculate the time
when our model disk's structure will be photoevaporation-dominated at
5~AU ($t_{\rm pe} < t_{\nu}$). Using Equation \ref{eq:tcrit} with
the photoevaporation profile ($\dot{\Sigma}_{\rm pe}(r)$) from
\cite{o12} appendix B (model 2 in \S~\ref{photoevap} of this work), we
find $t_\mathrm{crit}(r~=~5~\mathrm{AU})~=~7.3$~Myr, or 2.3~Myr after
$t_\mathrm{now}$. This is more than four times the viscous timescale at
5~AU, which suggests that giant planets currently forming by core
accretion in TW Hydrae's inner disk will still have a long epoch of
migration ahead, unimpeded by photoevaporation.

As a side note, \cite{i13} measure the accretion rate of TW Hydrae to be
$\dot{M} = 1.8 \times 10^{-9}~\mathrm{M_\sun yr^{-1}}$, which
corresponds to $\alpha$=0.0002 if we assume steady-state accretion.
However, the presence of a giant planet in the inner disk would greatly suppress accretion and \cite{a16} found evidence of a narrow
gap at 1~AU that could be formed by a giant planet. Our choice of
$\alpha$=0.001 is probably a good estimate for this system.

Our analytic calculations suggest that giant planets should have ample
time to migrate before photoevaporation can
begin to sculpt the disk surface density profile, modifying their migration rates.
We next verify our analytic results using numerical simulations with the FARGO code.


\section{FARGO simulations} \label{simulations}

Having demonstrated that newly formed giant planets should migrate for
several viscous timescales unimpeded by photoevaporation, we now assess
how planets migrate once 
photoevaporation begins to sculpt the disk surface density profile. Simulations
combining photoevaporation with giant planet migration have been performed
in 1-D \citep{aa09,ap12,er15}. We use the FARGO 2-D code \citep{m00} instead
of 1-D models because of the importance of non-axisymmetric flows in migration
\citep{p14}. Though we use an axisymmetric prescription for
photoevaporative mass loss $\dot{\Sigma}_{\rm pe}(r)$, the {\it relative}
surface density change induced by photoevaporation can be very high for
the non-axisymmetric tidal tails and co-rotating horseshoe in the
planet's tidal gap because of their low densities compared to the
surrounding disk.
Simulations of giant planets migrating in 2-D photoevaporating disks have been
performed using FARGO in previous studies \citep{ma12,r13x,r15}. \cite{ma12}
study how planet orbital distributions from planet-planet scattering evolve
during the gas disk phase and the n-body phase after the gaseous disk is
dispersed by photoevaporation. In contrast to our work, they focus on multi-planet
systems, where planet-planet resonance interactions tend to be more important
for migration than planet-disk interactions. \cite{r13x,r15} study how giant
planets inhibiting disk accretion across their orbits leads to various shapes
of transitional disks during photoevaporative clearing. To our knowledge, no
previous simulations have been carried out in 2-D which attempt to discover
how photoevaporation affects single giant-planet migration or the semimajor axis
distribution of giant exoplanets.

We divide our FARGO 2-D simulations into 3 categories: Planet-Forming
Disks, Fixed Orbits, and ER15 Comparisons/Extensions. In our Planet-Forming Disks
we try to answer the question, if a Jupiter-mass planet forms as late as
theoretically possible in a disk's lifetime, when the disk is depleted
beyond the minimum mass that will still support giant planet formation, 
will it still be migrating when photoevaporation can start to
sculpt the disk surface density profile? (\S~\ref{PFDsimulations}).
A second set of simulations with
fixed orbits allows for a high-resolution study of how photoevaporation
affects non-axisymmetric flows inside the planet's tidal gap
(\S~\ref{FOGSsimulations}). Finally, ER15 Comparison simulations use
similar parameters to the XEUV (X-ray + EUV photoevaporation) disk
models in ER15, allowing a direct comparison of our methods and results
to previous work (\S~\ref{1Dsimulations}).

\subsection{Planet-Forming Disks} \label{PFDsimulations}

We construct our Planet-Forming Disk simulations to reflect the
circumstances under which giant planets survive until photoevaporation
can sculpt surface density changes in the disk. Since giant planets
which form early in the lifetime of disks may not survive
\citep[e.g.][]{kn12}, we start our disks with a surface density profile,
\begin{equation}
\Sigma (r) = 500\ {\rm g~cm^{-2}} \left( \frac{\rm 1 AU}{r} \right),
\end{equation}
which is roughly 1/3 to 1/5 of estimates of the lowest disk density that can still form a
giant planet in the range of initial semimajor axis locations predicted
by the Nice model \citep{tg05,j07,l09}. In contrast to our analytic disk
models in \S \ref{timescales}, the disks we now label ``planet-forming''
have lost too much mass to keep forming giant planets by planetesimal
accretion, but should still be capable of forming Neptune analogs.
While our surface density in these models is lower than estimates of the
minimum surface density needed to form Jupiter analogs, it differs
significantly from the previous studies by AP12 and ER15 who consider
surface densities all the way down to those found at the ``disk clearing time" when
photoevaporation quickly disperses the remaining disk. \cite{aa09} explain that
for gas giants to survive in their models, the gas giants must form in disks with
surface densities at 5-10 AU of $\la 10$~g~cm$^{-2}$, or $\sim$1.5 orders of
magnitude less than predictions of the minimum densities required \citep{j07,l09}.
The semimajor axis distribution features predicted by AP12 and ER15 are
enhanced by these late-forming Jupiters, which may not have any
physical analogs. However, since all of the planets in AP12 and ER15 start at
5~AU, these late-forming Jupiters may be representative of planets that form
much further out in the disk and migrate inward, reaching $\sim$5 AU as the
disk is about to disperse. Pebble accretion may allow giant planets to form at much
larger distances from the star than traditional planetesimal accretion
\citep{lj12,mn12,l15,b15a}, but pebble accretion models have so far
focused on disks with $\Sigma_{\rm 1 AU} = 1700$~g~cm$^{-2}$,
more than three times as dense as our model disk. It seems possible that a
combination of giant planet migration and viscous evolution of the disk
may bring some giant planets formed by pebble accretion to $\sim$5~AU at
late times. To allow a more direct comparison with AP12 and ER15,
we simulate recently-formed Jupiters in much lower disk masses
(\S~\ref{1Dsimulations}), and discuss the results of those simulations in
\S~\ref{1Dresults}.

In our Planet-Forming Disk simulations, we assume a blackbody
temperature profile for our disks using a luminosity
of $0.63 L_{\sun}$, corresponding to a solar-mass star aged 5~Myr
\citep{dm94}. This age is appropriate for a disk near in time to
dispersal by photoevaporation \citep[$\sim$6 Myr in][]{a06b}, yet
massive enough not to be very far past the giant planet formation epoch
(see \S~\ref{timescales}).  For turbulent viscosity we consider
$\alpha=0.001$ and $\alpha=0.01$, but devote much more computational
time to $\alpha=0.01$ because it results in faster accretion, allowing
the simulations to reach lower disk densities and greater photoevaporative sculpting. The physical effects of the $\alpha$
parameter are discussed more in \S~\ref{FOGSresults}.

For the simulations with $\alpha = 0.01$, each disk contains one
Jupiter-mass (0.001 $M_\sun$) or Neptune-mass (0.00005 $M_\sun$) planet,
and the accretion rate onto the planet is set to zero.
We note, however, that accretion onto the planet can both alter the rate
of migration by removing material from near the planet's orbit, and reduce the
mass transfer efficiency into the inner disk, triggering the UV-switch
\citep{d02,b03,aa09,wc10,ap12,r13x,dk17}.
For each planet mass, we test each of the three
photoevaporation models described in \S~\ref{photoevap}. For each
photoevaporation model (EUV, X-ray, or FUV) and each planet mass
(Jupiter or Neptune), we consider three starting locations: $a_{\rm
start} = (2/3) r_{\rm pe}$, $a_{\rm start} = r_{\rm pe}$, and $a_{\rm start} =
(4/3) r_{\rm pe}$, where $r_{\rm pe}$ is the disk radius with the minimum
photoevaporation timescale, given by ${\rm min}(\Sigma(r, t=0) /
\dot{\Sigma}_{\rm pe}(r))$. We chose the three planet starting locations to
capture the effects of (1) having only the planet's outer Lindblad
resonances in the photoevaporating gap, (2) all of the Lindblad
resonances in the gap, and (3) only the inner Lindblad resonances in
the gap. We also simulated the migration of Jupiter-mass planets in
disks with $\alpha$=0.001, using the X-ray photoevaporation (model 2)
with each corresponding value of $a_{\rm start}$. A summary of all
simulation parameters can be found in Appendix~\ref{AppendixParams},
Table~\ref{params}.

For each planet mass, photoevaporation model, value of $\alpha$, and
$a_{\rm start}$, we run a control simulation with the same planet mass,
$\alpha$, and $a_{\rm start}$ but no photoevaporation (42 simulations in
total).  We can then separate the effects of photoevaporation from the
effects of other parameter choices. To mitigate the numerical effects of
suddenly placing a giant planet into a disk, while still restricting the
scope of our simulations to Type II migration, we first run each
simulation for about 300 orbits with the planet's migration turned off,
then release the planet. This allows the system to stabilize and the
planet's tidal gap to form before it is allowed to migrate.  Preliminary
runs varying radial and azimuthal grid resolution suggested the code was
stable and accurate for the chosen grid of 600 radial by 200 azimuthal
zones spanning $\sim$1.7--200~AU for X-ray and FUV photoevaporation. We
require 800 radial and 200 azimuthal zones spanning $\sim$0.3--200~AU
for EUV photoevaporation, which forms a gap between 1 and 2~AU. The
radial grid is equally spaced in $\log(r)$, which provides ${\sim}$9 zones in a Jupiter-mass planet's Hill radius, sufficient to resolve the
corotation region \citep{m02,pp09}.

Finally, we ran test simulations for Jupiter-mass planets in disks with
both $\alpha = 0.01$ and $\alpha = 0.001$, this time adding an
exponential taper on the migration torque within the planet's Hill
sphere. While the value of $\alpha$ and the inclusion of an exponential
Hill torque taper did significantly alter the planet migration rates,
they did not contradict our results that migration tracks in disks that
are massive enough to form giant planets are mostly unchanged by
photoevaporation (see discussion of results in \S~\ref{PFDresults}). For
replication convenience, Table~\ref{PFDparams} gives our Planet-Forming
Disk model parameters in the format of a FARGO parameter file.

\subsection{Fixed Orbits} \label{FOGSsimulations}

Since the planet's tidal gap has lower surface density than the
surrounding disk, the material within has some of the lowest
photoevaporation timescales. The Fixed Orbits parameter study consists
of high-resolution simulations of photoevaporation's effect on the gas
directly surrounding the planet. As the tidal gap structure may change
with time due to both migration and photoevaporation, keeping the
planet's orbit fixed allows us to isolate the impact of photoevaporation
on the gap structure. Fixing the planet's orbital radius also permits a
closer-in outer boundary, improving resolution and computational speed.
We can therefore carry out a precise study of how planets' tidal gaps
are affected by photoevaporation without simulating the planets' entire
migration paths. We directly compare simulations where photoevaporation
is on and off to quantify how photoevaporation affects the density
within the planet's tidal gap and hence the migration torques.

Here we use the same set of parameters as in our Planet-Forming Disk
model except for the disk outer radii, which are reduced to $\sim$60~AU
for FUV/X-ray and $\sim$10~AU for EUV photoevaporation simulations to
give us a higher resolution in the tidal gap. We leave the aspect ratio,
$\Sigma(r)$, $\dot{\Sigma}_{\rm pe}(r)$ profiles, and planet starting locations
unchanged (see \S~\ref{PFDsimulations}). We include only Jupiter-mass
planets since Neptune-mass planets do not carve tidal gaps. In order to
study photoevaporation's effect on the corotation torque---which can
switch the migration direction from inward to outward for high-viscosity
disks \citep{cm07}---we run each simulation using both $\alpha = 0.001$
and $\alpha = 0.01$.
In the $\alpha = 0.01$ disks, the higher viscosity makes it difficult for the
planet to carve a deep tidal gap, resulting in $\sim$30 times higher surface
density in the gap compared to the $\alpha=0.001$ disks \citep{f14,dk15}.
Here the higher density in the tidal gap can generate corotation torques that can
significantly alter the planet's migration rate (unlike in the $\alpha = 0.001$
disks), but the tidal gap density is still
much lower than in the rest of the disk.

Besides resolution, the only difference between these simulations
and the Planet-Forming Disks is here we are keeping every planet's orbit
fixed: we update the disk density based on torque from the planet, and
we calculate the migration torques but do not apply them to the planet.
Once again, every photoevaporating disk is paired with a photoevaporation-off
control simulation. We compare these simulations after 50~kyr, which is
sufficient time for the planets' tidal gaps to form, and for photoevaporation
to deplete $\sim$10\% of the disk mass in our FUV models, which have
the greatest mass-loss rate. See Appendix~\ref{AppendixParams} for specific simulation parameters.
We note that keeping the planets' orbits fixed does not conserve angular
momentum, but the cumulative error this introduces is small in
the simulations we present here. For example, in our
disks undergoing FUV photoevaporation, the average torque the planet exerts
on the disk is $\bar{\tau} \approx 5 \times 10^{-8}$~M$_\star$~AU$^2$~yr$^{-2}$,
but the disks start out with a total angular momentum of
$L \approx$ 0.13~M$_\star$~AU$^2$~yr$^{-1}$, so the error in angular momentum after 50~kyr ($\Delta L / L = \bar{\tau} \Delta t / L$) is roughly $2\%$.

\subsection{ER15 Comparisons and Extensions} \label{1Dsimulations}

Previous numerical studies of how photoevaporation affects the semimajor
axis distribution of exoplanets (AP12 and ER15) used 1-D population
synthesis models. These models evolve disks using the 1-D viscous
evolution equation for thin disks \citep{p81}, a 1-D prescription for
planet migration \citep{lp86,a02}, and a photoevaporation term
$\dot{\Sigma}_{\rm pe}$ \citep{c01}. The computational resources saved by
moving from 2-D to 1-D disk models allow exploration of a larger
parameter space. While AP12 and ER15 both include disks with surface
densities comparable to our Planet-Forming Disks in their simulation
sets, their analysis is statistical: they do not map individual outcomes
to unique parameter combinations,
making it difficult to see which part of their simulation parameter space
generates their deserts and pileups. However,
their models are very similar to previous 1-D models of giant
planet migration in photoevaporating disks, which find that planets can only
survive if they form in the last 10-20\% of the disk lifetime \citep{a02,a07,aa09}.
Hence it seems likely that the effects observed by AP12 and ER15 originated
from lower disk masses at the time of planet formation, which we did not
explore in our Planet-Forming Disk simulations (\S~\ref{PFDsimulations}).

Our ER15 Comparisons are FARGO 2-D simulations with the same disk
viscosity, aspect ratio, $\Sigma(r)$ profile, $\dot{\Sigma}_{\rm pe}(r)$
profile, and planet starting locations as the ``XEUV" disk models of
ER15. Due to computational time constraints, we only simulate 3 planet
masses (0.5, 1, and 2 $M_{\rm J}$), while ER15 draw their planet masses
from a 0.5-5 $M_{\rm J}$ uniform distribution. Unlike ER15, who chose
migration start times from a uniform distribution between 0.25~Myr and
the disk-clearing time, we start all planets at the same time in disk
evolution. We choose our start time so the disk has 1/10$^{\rm th}$ of its original
mass ($0.007 M_\sun$ out of the initial $0.07 M_\sun$ in ER15), so
photoevaporation will start to affect the surface density profile of the
disk during each simulation. Using ER15's viscous timescale of $1.2
\times 10^6$~yr at 10~AU, we calculate $\alpha=0.0009545$ for our
simulations by assuming a blackbody heating temperature profile (see
\S~\ref{timescales}) for the luminosity, $L=0.17 \rm L_\sun$ appropriate
for a $0.7 M_\sun$ star \citep{dm94}. For simplicity we will refer to
$\alpha=0.0009545$ as the rounded value $\alpha=0.001$ in this paper. We
use a 600 radial by 200 azimuthal zone grid with a 1~AU inner boundary.
We also ran three test models with all parameters held equal except for
using 800, 1000, and 1200 radial zones and a smaller inner boundary
(0.3~AU) for 10~kyr for comparison, and found no significant differences. Planetary
accretion is neglected as the planet is again assumed to have reached
nearly terminal mass.
(Note that both
AP12 and ER15 find that accreting planets can reduce the mass flux
through the gap, draining the inner disk, triggering the UV-switch, and
halting their own migration by accelerating disk dissipation; we did not
explore this scenario).
The ER15 Extensions models are extensions of the ER15 Comparison
parameter space to include high viscosity ($\alpha$=0.009545, hereafter
referred to as the rounded value 0.01) and lower disk mass (reduced by
another factor of 10 to $0.0007 M_\sun$). High-viscosity models allow a
more direct comparison to our Planet-Forming Disk migration tracks, and
lower disk mass shows what happens in the extreme case when the disk is
tenuous enough for photoevaporation to clear it in much less than the
typical Type II migration timescale of ${\sim}$10$^5$~years. As in the
Planet-Forming Disk models (\S~\ref{PFDsimulations}), each
photoevaporating disk is paired with a control simulation where
photoevaporation is turned off. See Appendix~\ref{AppendixParams} for
specific simulation parameters.

In the next section, we describe the results of each simulation set,
focusing specifically on the differences between the photoevaporating
disks and the control simulations with photoevaporation turned off.

\section{Results} \label{results}

Our simulations consistently indicate that photoevaporation has little
effect on planet migration. In \S~\ref{PFDresults}, we confirm previous
results showing that disks with enough mass to form giant planets have
strong migration torques that are minimally affected by photoevaporation
\citep{hi13}, consistent with planet-forming disks being
accretion-dominated as defined in \S~\ref{timescales}. In
\S~\ref{FOGSresults}, we present detailed calculations of migration
torques, with and without photoevaporation, from our fixed-orbit models.
Finally, in \S~\ref{1Dresults} we demonstrate that even within the same
parameter space of disk mass, viscosity, and photoevaporation model
simulated by ER15,
the only significant impact photoevaporation has on planet
migration is to halt very slowly migrating planets by dissipating the gas disk.

\subsection{Planet-Forming Disks Results} \label{PFDresults}

\begin{figure}
\epsscale{0.6}
\plotone{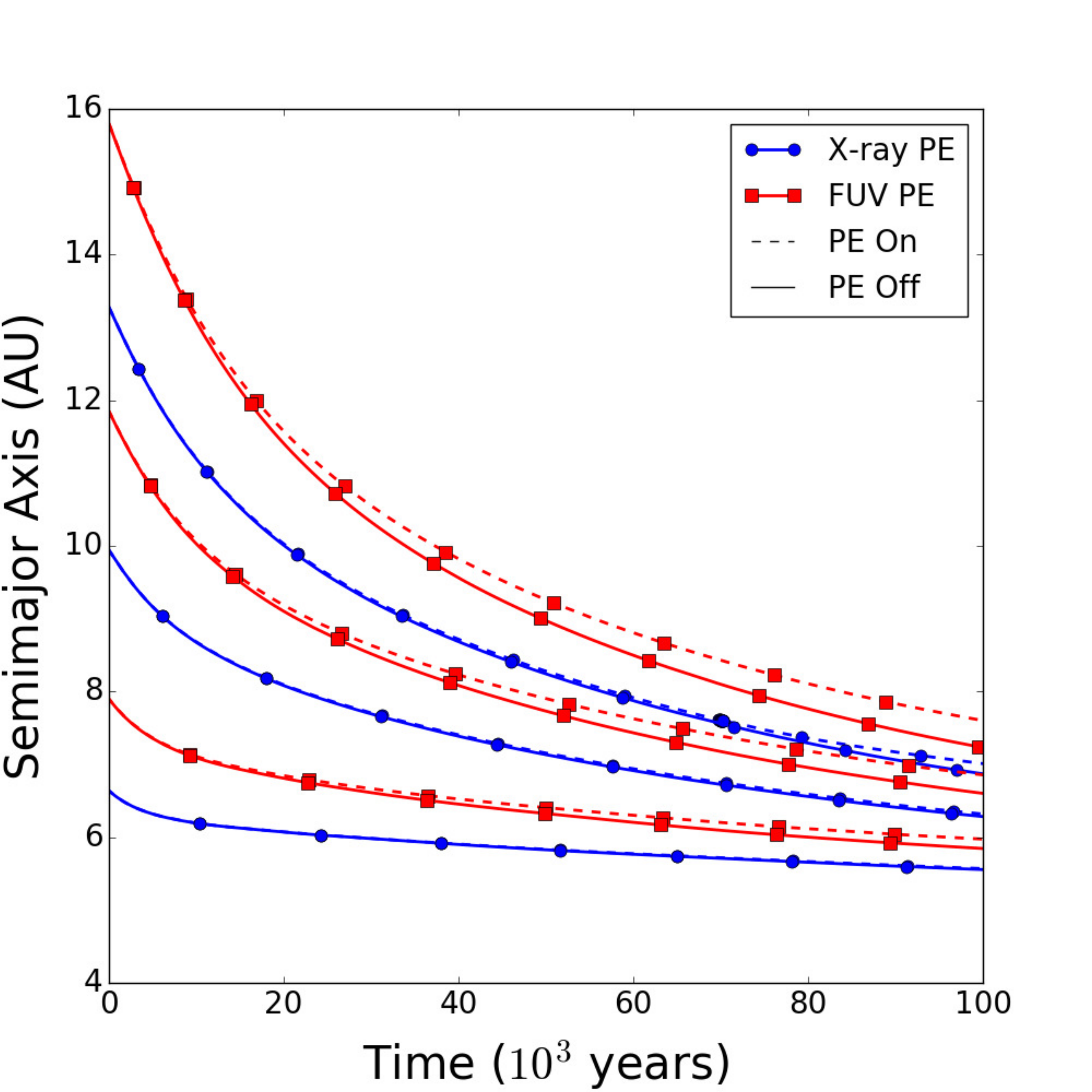}
\caption{A comparison of migration tracks of Jupiter-mass planets in our
Planet-Forming Disk models, showing X-ray and FUV photoevaporation (PE)
mass-loss profiles plus control simulations with photoevaporation off.
The initial planet positions for each model are 2/3, 1, and 4/3 of each
photoevaporation model's gap-opening radius, found by minimizing $t_{\rm
pe} = \Sigma(r) / \dot{\Sigma}(r)$. \label{fig2}}
\end{figure}

\begin{figure}
\epsscale{0.6}
\plotone{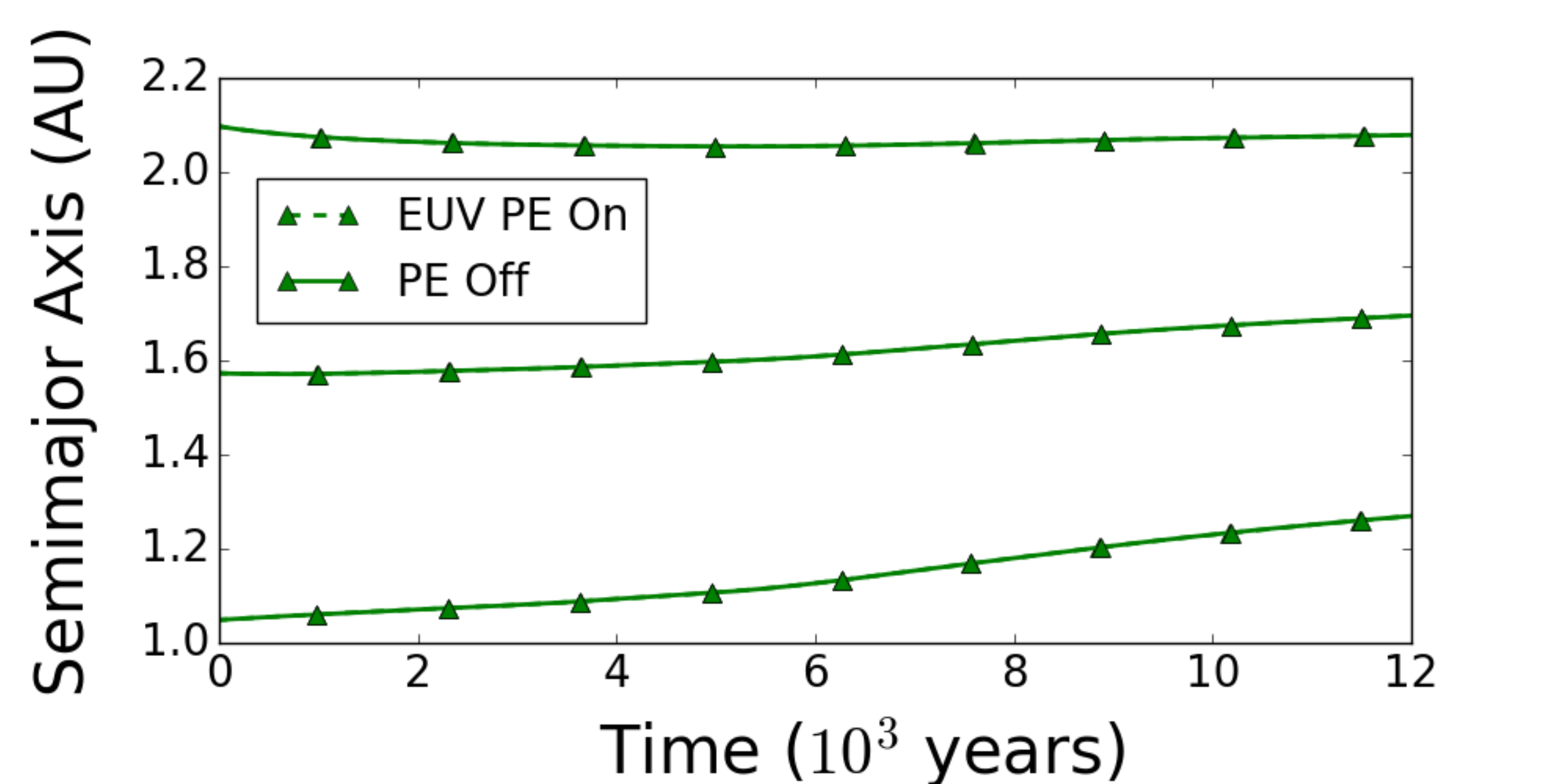}
\caption{A comparison of migration tracks of Jupiter-mass planets in our Planet-Forming Disk models, showing the EUV photoevaporation (PE) mass-loss profile plus control simulations with photoevaporation off. The initial planet positions for each model are 2/3, 1, and 4/3 of each photoevaporation model's gap-opening radius. The PE-on migration tracks are identical to the PE-off migration tracks. \label{fig3}}
\end{figure}

Here we explore the effects of EUV, X-ray, and FUV photoevaporation
\citep[][see \S~\ref{photoevap}]{f04,gh09,o11} on Type II migration in
disks recently capable of forming giant planets.  Figures~\ref{fig2} and
\ref{fig3} show migration tracks for planets placed just interior to the
nascent photoevaporated gap, in the middle of the gap, and just exterior
to the gap, for all three $\dot{\Sigma}_{\rm pe}(r)$ profiles (dashed lines) and a
disk with $\alpha = 0.01$. Control simulations are conducted with no
photoevaporation (solid lines). The migration tracks for planets in
disks with and without photoevaporation are strikingly similar for the
X-ray and FUV models (Figure~\ref{fig2}); for the EUV case
(Figure~\ref{fig3}), the migration tracks in the photoevaporation on/off
cases are identical. Despite the planets being placed near the location
where the photoevaporation timescale $t_{\rm pe}$ is shortest, there is
no indication that
migration rates are significantly slowed due to a widening
photoevaporation-induced gap.
Instead, the photoevaporative mass loss merely lowers the migration
torque slightly, slowing the planet so that its semimajor axis after
0.1~Myr of migration is a maximum of 5\% higher than it would be in a
non-photoevaporating disk. The only notable difference between the three
$\dot{\Sigma}_{\rm pe}(r)$ profiles is the total predicted disk mass-loss rate,
where a higher photoevaporation mass-loss rate (FUV model $>$ X-ray model $>$ EUV model)
results in more slowing of migration due to disk depletion---though
effects are almost negligible in all cases. Our results demonstrate that
Jupiter-mass planets will have a period of migration unaffected by
photoevaporation that lasts at least 0.1~Myr after their formation,
consistent with the results of \cite{hi13}.

\begin{figure}
\epsscale{0.8}
\plotone{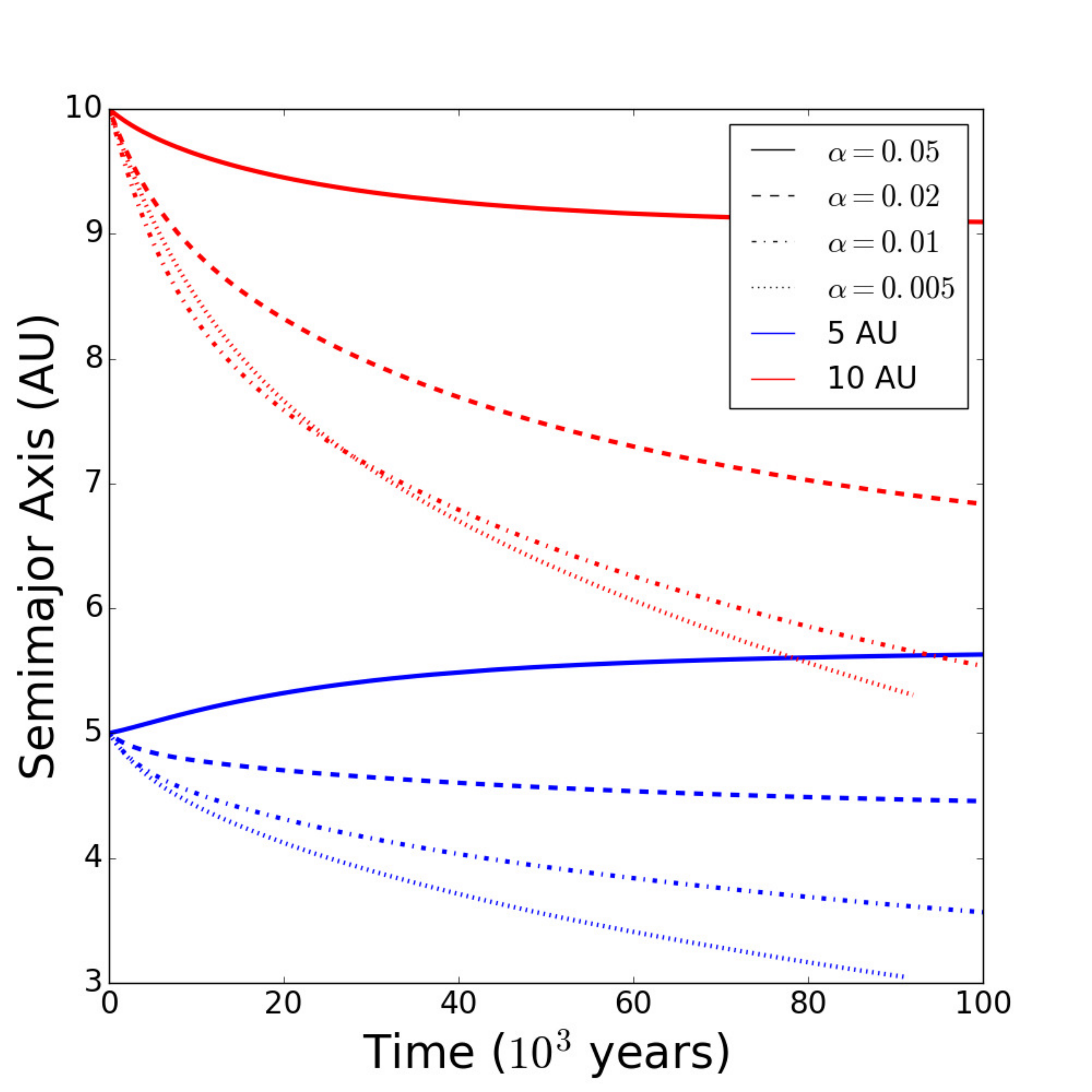}
\caption{Migration tracks of 1.0 $\mathrm{M_{J}}$ planets starting at 5
and 10~AU, in disks of varying viscosity. Disk parameters are nearly
identical to \cite{cm07} \S~3 which places planets at 5~AU. We add a set
of planets 10~AU to show how corotation torque is stronger at smaller
radii. Photoevaporation is not included. Our comparisons with
the \citet{cm07} models show that disk viscosity and migration
starting location are the dominant parameters that determine the
migration tracks shown in Figure~\ref{fig2}. \label{fig4}}
\end{figure}

Despite the barely noticeable effect photoevaporation has on our
migration tracks, the planets in our X-ray and FUV simulations (and
their control simulations with no photoevaporation) all seem to converge
at 6-8 AU. 
This is due to the corotation torque, which tends to push planets outward, and grows
as planet mass and disk viscosity increase and as disk radius
decreases \citep{m01}. The relationships between disk viscosity, planet
mass, and migration torque have been studied using disk simulations in
1-D, 2-D, and 3-D \citep[e.g.][]{cm07,m07,b13a}. Since we use 2-D disk
simulations like \cite{cm07}, we can compare our results to theirs
directly and extend the parameter space of their simulations. In
Figure~\ref{fig4} we present nearly identical simulations to those of
\cite{cm07} featuring planets migrating after formation at 5 AU in blue
(see Appendix~\ref{AppendixParams} for FARGO parameters).
We extend the work of \cite{cm07}
by adding a second set of migration tracks for planets formed at 10 AU
in red.  Figure~\ref{fig4} shows that two equal-mass planets in a disk
with uniform $\alpha$ turbulent efficiency can migrate in different
directions, or in the same direction at different rates, depending on
their initial orbital radii, which accounts for the planets placed at
the outside of each photoevaporating region in Figure~\ref{fig2}
``catching up'' to the planets starting at smaller semimajor axes.
Likewise, planets starting at the same location can migrate either
inward or outward depending the strength of the corotation torque
\citep{bm08,pp08,mc09,kl12,p12}. We find that the disk viscosity has a
far stronger effect on migration tracks than photoevaporation in a disk
with enough mass to have recently formed planets. Indeed, a set of
simulations identical to those in Figure~\ref{fig2}, except with $\alpha
= 0.001$ instead of $\alpha = 0.01$, finds much slower migration but
still no substantial effects caused by photoevaporation
(Appendix~\ref{AppendixMigration}, Figure~\ref{fig15}).

\begin{figure}
\epsscale{0.8}
\plotone{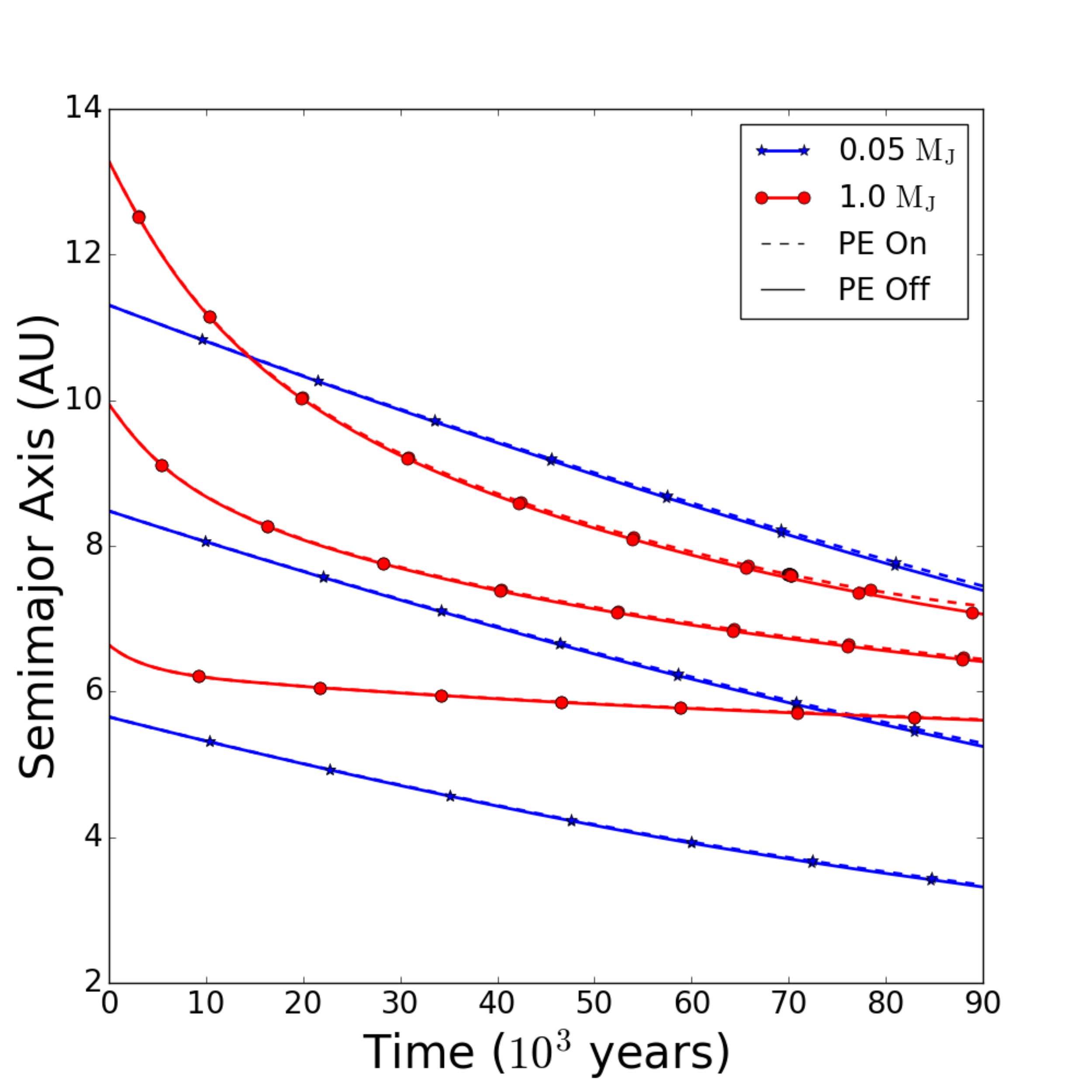}
\caption{Migration tracks of Neptune (0.05 M$_\mathrm{J}$) and
Jupiter-mass planets with starting locations near where X-ray
photoevaporation (PE) opens its gap. \label{fig5}}
\end{figure}

Planet mass also plays an important role in determining migration tracks
\citep{w97,m02,kc08,k09,bk11b}. Figure~\ref{fig5} compares migration
tracks of Neptune-mass planets and Jupiter-mass planets that start from
the same orbital radii in the disk with $\alpha = 0.01$ (same as in
Figure~\ref{fig2}). Dashed lines show planets in disks being
photoevaporated by X-ray radiation (model 2 in \S~\ref{photoevap}), and
solid lines show control simulations with no photoevaporation. The
Neptune-mass planets are not massive enough to open a gap in the disk,
so they experience Type I migration. Here, too, photoevaporation has a
barely discernible effect on migration tracks. Variations in corotation
torque are clearly the dominant force in shaping migration tracks in
our Planet-Forming Disks, and we must conclude that the impact of
photoevaporation on migration is negligible in the era directly
following giant planet formation.

\clearpage

\subsection{Fixed Orbits Results} \label{FOGSresults}

In \S~\ref{PFDresults} we showed that in a disk with $\Sigma = \left(500
\; {\rm g \; cm^{-2}} \right) \left( 1 {\rm AU}/R \right)$,
roughly 1/3 to 1/5 of the minimum surface density that allows Jupiter-mass
planets to form near 5~AU \citep{j07,l09}, photoevaporation has almost no
effect on planet migration. Instead, migration speed and direction are
primarily determined by the corotation torque, which in turn is a
function of viscosity and planet mass \citep{bm08,pp08}.
More recent 3-D
simulations of accreting gas giants have shown that the corotation
torque is much higher than predicted in 2-D, and may slow predicted
migration rates by a factor of 3 \citep{faw15}.
To discover how
photoevaporation may modify corotation torque, we have carried out the
set of simulations described in \S~\ref{FOGSsimulations}, where we allow
the planet to torque the disk and modify the disk's density structure,
but do not allow the disk to torque the planet. However, though we hold
the planet on a fixed orbit, we calculate the torque the disk would
exert on the planet at select time snapshots and pinpoint the locations
where photoevaporation is modifying the migration torque.
Even though we carried out identical Fixed Orbits simulations for all three
photoevaportion models listed in \S~\ref{photoevap}, we only present detailed
results for the FUV photoevaporation model as it has the greatest
photoevaporative mass-loss rate, making it an ideal candidate for
visualizing how photoevaporation affects tidal gap structure.
FUV-driven mass loss is mostly external to the planet's orbit, but
we find that the FUV photoevaporation model still causes greater
depletion of the tidal gap after 50,000 years than either of the EUV or
X-ray photoevaporation models. Outside the tidal gap, viscous forces
dominate the gas surface density profile, so the photoevaporative
mass-loss rate affects the surface density but the photoevaporation
profile shape $\dot{\Sigma}(r)$ does not.
The effects of FUV photoevaporation on tidal gap
structure that we present here should be regarded simply as scaled-up
versions of the effects of EUV and X-ray photoevaporation.

\begin{figure}
\centering
\begin{tabular}{c}
\includegraphics[scale=0.42]{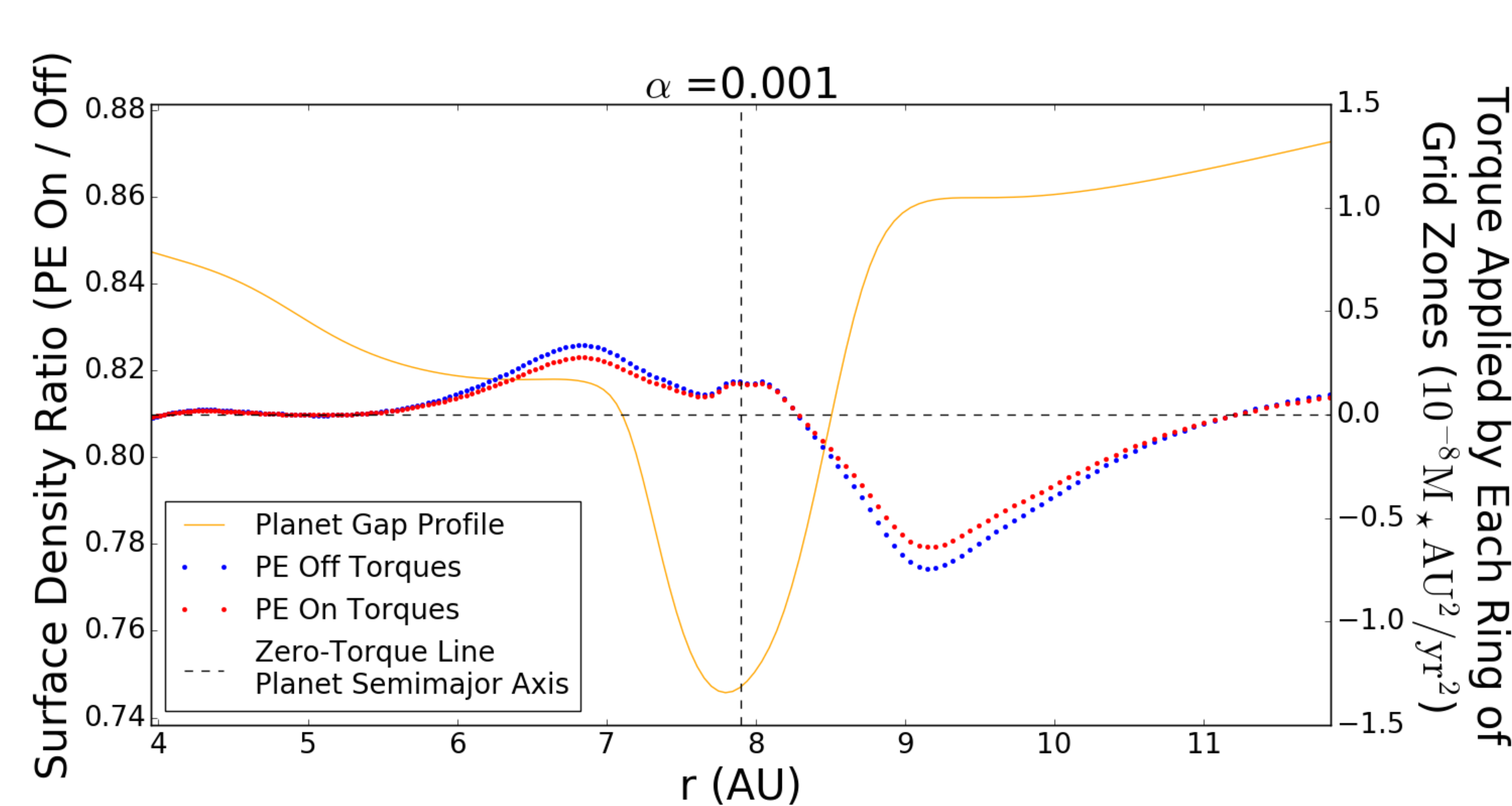} \\
\includegraphics[scale=0.42]{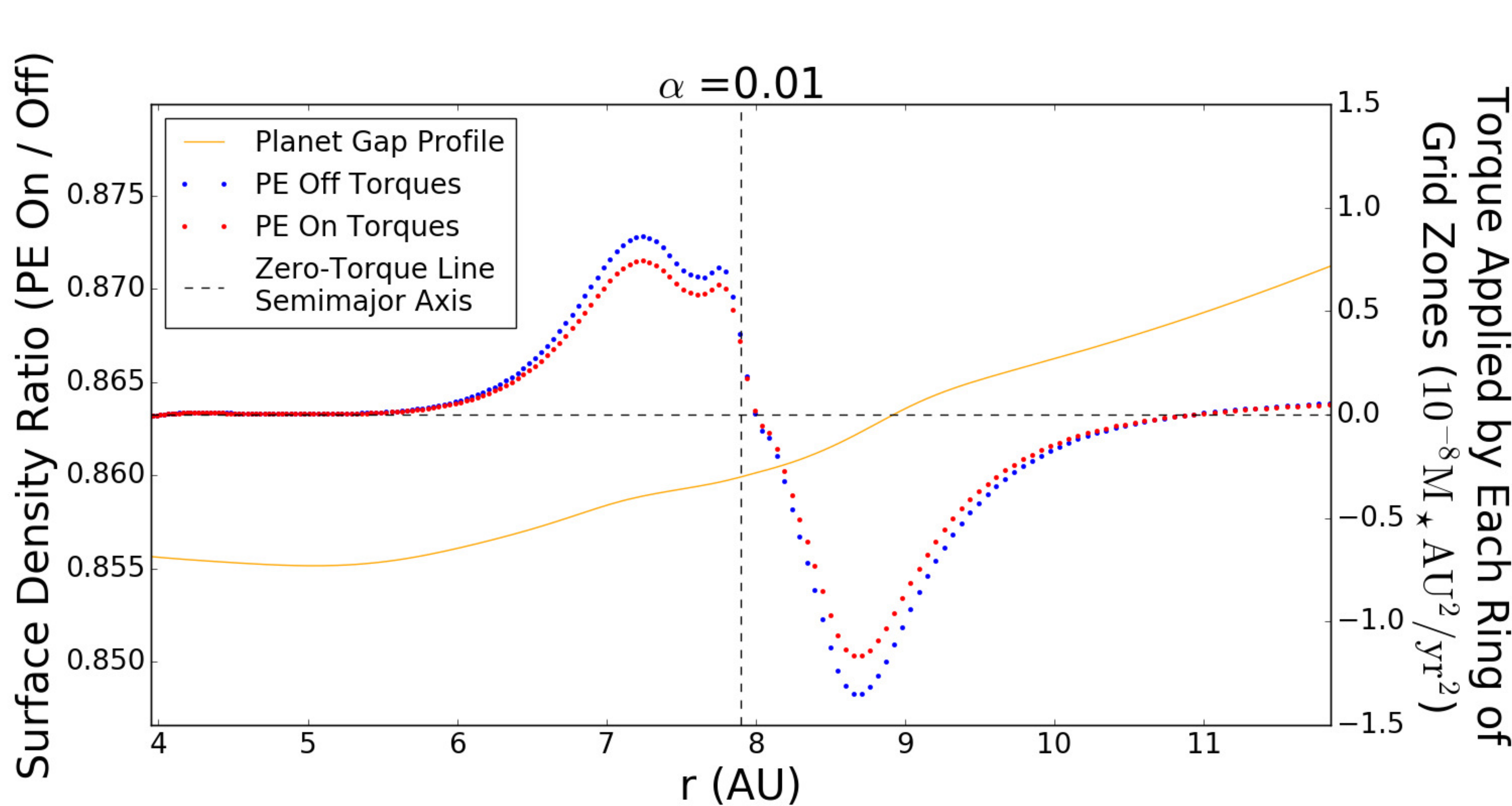}
\end{tabular}
\caption{Torque per ring of grid zones and relative density profiles for Fixed Orbit Simulations at 8 AU using the FUV photoevaporation (PE) profile (greatest mass-loss rate), after 50 kyr. Note that because we are plotting the torque per ring of grid zones, the total torque is not the area under the dotted curve, but rather the sum of the individual dots in the curve. Both $\alpha=0.01$ and $\alpha=0.001$ show $\sim$13\% disk mass depletion due to photoevaporation. Only $\alpha=0.001$ shows photoevaporation deepening the planet's tidal gap, as the lower viscosity at the gap, which has the lowest density of any point in the disk, lets the viscous timescale $t_\nu$ exceed the photoevaporation timescale $t_{\rm pe}$. \label{fig6}}
\end{figure}

In Figure~\ref{fig6}, we show migration torques after 50,000~years of
disk evolution and FUV photoevaporation
in disks with $\alpha = 0.001$ (top) and $\alpha = 0.01$
(bottom, same as in \S~\ref{PFDresults}). We fix
the planet orbits at 8~AU, since a planet starting at $r_{\rm pe} =
12$~AU migrates about 4~AU in 50,000~years in this model disk
(Figure~\ref{fig2}). 
We show migration torques in the photoevaporating (red) and
non-photoevaporating (blue) disks. In orange, we over-plot $\Sigma_{\rm
pe}(r) / \Sigma(r)$, the ratio of surface density in the
photoevaporating disk to surface density in the control,
non-photoevaporating disk. Values below unity indicate areas depleted by
photoevaporation. Note that the tidal gap opened by the planet is
present in simulations both with PE on and PE off, so it will not show
up in Figure~\ref{fig6} {\it unless} deepened by photoevaporation, as
seen in the disk with $\alpha = 0.001$. For both values of $\alpha$,
photoevaporation depletes the global disk density by $\sim$13\% after
50,000~years, reducing migration torque per annulus by roughly the same
percentage. The extra depletion in the planet's tidal gap in the disk
with $\alpha = 0.001$ hardly alters the net migration torque, as the
strongest torque comes from outside the tidal gap (recall that the
corotation torque weakens in weakly turbulent disks; see
\S~\ref{PFDresults} and Figure~\ref{fig4}). The fact that
photoevaporation depletes the disk globally, instead of only in a narrow
annulus, results from the photoevaporation timescale exceeding the
viscous timescale throughout the disk, so viscous transport of disk
material refills depleted regions faster than they can be carved out by
photoevaporation. Migration torques are reduced over broad regions of
the disk, but in a manner indistinguishable from other large-scale
sources of disk depletion such as accretion onto the star---no
surface density gradients steep enough to significantly alter migration tracks are formed as a result of photoevaporation.

We see a small localized effect in the planet's tidal gap in
the disk with $\alpha = 0.001$, where the low disk density and long
viscous timescale allow photoevaporation to deplete a narrow region by a
further 10\% on top of the overall $\sim$13\% depletion in the entire
plotted region. This depletion is removing material that would
provide a positive (outward) torque, so the planet's inward migration
would speed up if we applied migration torques in this simulation set.
However, after a short epoch of fast inward migration, the planet would
move interior to the depleted annulus and its migration rate
would slow down. A density reduction in the corotation region would
create a burst of migration speed that forces the planet to quickly move
into a less depleted region of the disk, becoming a self-limiting
process: the deeper the depletion at corotation, the faster the planet
moves into a denser region. Indeed, in Figure~\ref{fig8} in the next
section, we see a slight initial speed-up in migration of the 1.0
M$_{\rm J}$ planet in an $\alpha=0.001$ disk as photoevaporation has had
a chance to clear out co-rotating material before we turn on the
migration torques and ``release'' the planet (see \S~\ref{simulations}),
but the long-term behavior is dominated by a slight slowing of migration
due to global disk depletion.

\begin{figure}
\epsscale{1.0}
\plotone{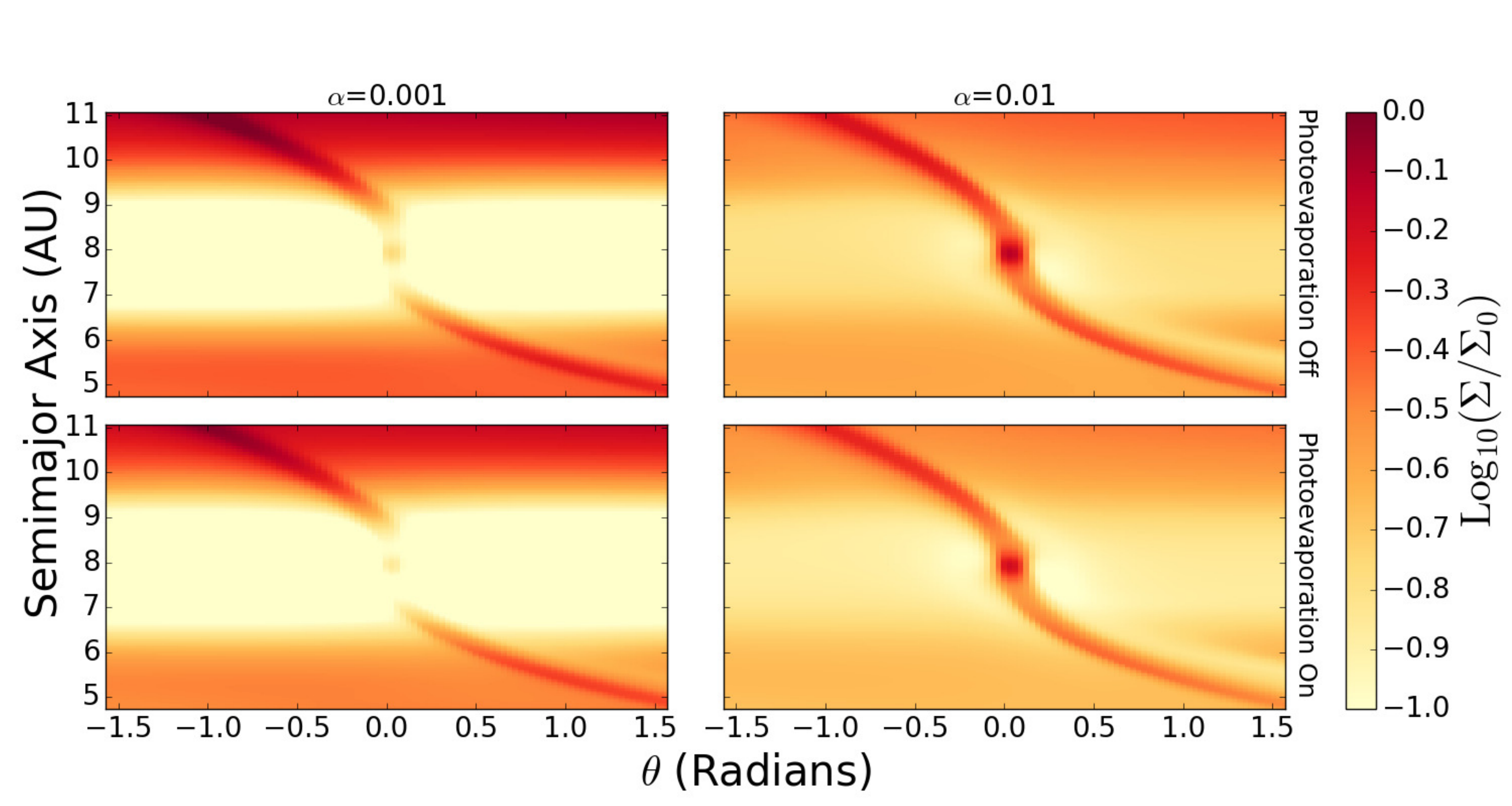}
\caption{Surface densities in the tidal gaps of Jupiter-mass planets in
fixed orbits at 8 AU under FUV photoevaporation (greatest mass-loss rate
model), after 50 kyr. The disk has an overall density reduction due to
photoevaporation, but the functional form of the density distribution in
the tidal gap remains essentially unchanged. \label{fig7}}
\end{figure}

Two caveats about interpreting the azimuthally averaged surface density
ratio, $\Sigma_{\rm pe}(r) / \Sigma(r)$, in Figure~\ref{fig6} are (1)
azimuthal averaging masks how photoevaporation is affecting
non-axisymmetric tidal flows and (2) the $\Sigma_{\rm pe}(r)/\Sigma(r)$
density ratio does not, by itself, show the planet's tidal gap
structure. In Figure~\ref{fig7}, we show pseudocolor plots of the tidal
gap structure normalized to the $t=0$, unperturbed surface density
profile, $\Sigma(r,\theta,t) / \Sigma(r,\theta,0)$, for $t =
50000$~years. The left column shows disks with $\alpha = 0.01$ and the
right-hand column shows disks with $\alpha = 0.001$; disks in the top
row experience no photoevaporation and disks in the bottom row are
photoevaporating according to model 3 (FUV). Even with the extra
$\sim$10\% photoevaporation-induced surface density reduction in the
tidal gap for the disk with $\alpha = 0.001$, the functional form of the
mass distribution inside the gap is relatively unaffected by
photoevaporation. More severe changes to the mass distribution inside
the gap may appear as the disk evolves beyond the 50,000~years of
photoevaporation simulated here, but it seems that a disk that recently
formed giant planets is massive enough, and the nascent photoevaporating
gap is wide enough, that a 10-20\% surface density depletion inside a
giant planet's tidal gap barely alters the migration torque balance. Our
detailed torque analysis from the Fixed Orbit Simulations confirms our
conclusion that newly-formed giant planets have a long migration epoch
where they experience very little interference from photoevaporation.

\clearpage

\subsection{ER15 Comparisons and Extensions Results} \label{1Dresults}

So far we have presented two sets of simulations in which
photoevaporation has little effect on giant planet migration. Our
results seem to 
point to a different conclusion from
AP12 and ER15, who
concluded
that photoevaporation sculpts
the semimajor axis distribution of giant planets.
However, until now we have only considered planets that are
formed in disks with significantly higher surface densities than the
AP12 and ER15 disks that host surviving planets.
Now we
analyze whether or not photoevaporation can significantly
affect giant planet migration tracks
using a simulation set with disk parameters following ER15.
To see how quickly planets may migrate near the disk clearing time,
we also include
extremely low-mass disks, in which photoevaporation creates strong
surface density gradients much more quickly than the planets migrate.
We also include
disks with $\alpha = 0.01$ for comparison with our Planet-Forming Disk
simulations (\S~\ref{PFDresults}).

\begin{figure}
\epsscale{0.9}
\plotone{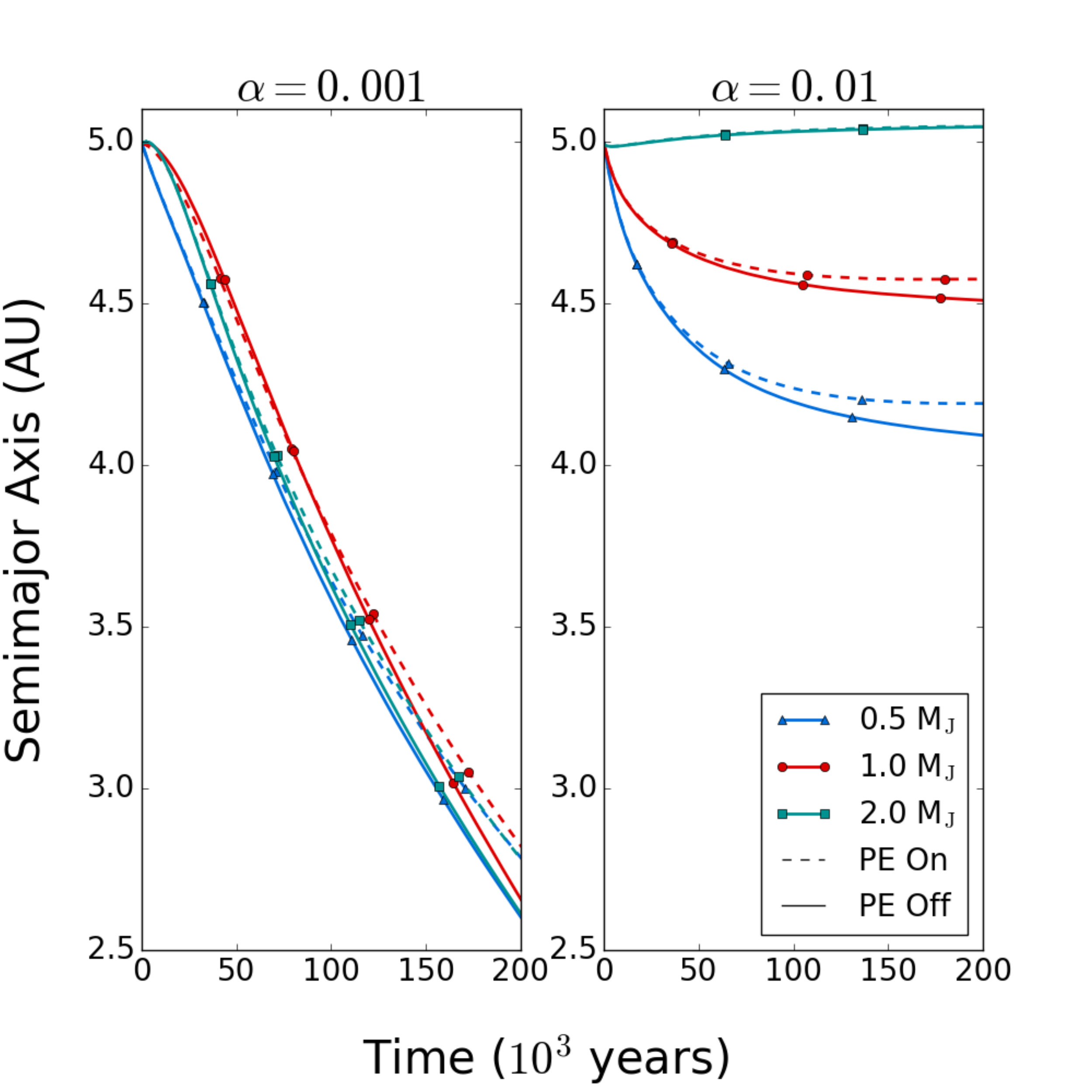}
\caption{Migration tracks for ER15 Comparison models with $M_\mathrm{disk}$=0.007 M$_\sun$. Note the PE on and off curves have roughly the same shape, indicating the lack of 
any significant perturbations to migration due to photoevaporation. \label{fig8}}
\end{figure}

Figure~\ref{fig8} (left panel) shows our ER15 Comparison model featuring
different planet masses.  As in the Planet-Forming Disks
(\S~\ref{PFDresults}), we find that including photoevaporation changes
the planets' semimajor axes by at most 3\% over the course of 200,000
years of migration. The left panel of Figure~\ref{fig8} demonstrates
that the ER15 models' lower viscosity results in less material in the
gaps, weaker corotation torque, and hence faster migration than in the
X-ray photoevaporating disks with $\alpha = 0.01$ plotted in
Figure~\ref{fig2}.  We extend the ER15 Comparison parameter space
to high viscosity in Figure~\ref{fig8} (right panel); all
other disk parameters remain identical to those of ER15 (except
we use 1/10$^{\rm th}$ their initial disk mass
to better match the conditions of their surviving planet population).
The dependence of
corotation torque on planetary mass \citep[e.g.][]{cm07} is evident in
the high-$\alpha$ migration tracks as here there is enough material
filling in the tidal gaps for corotation torque to dominate migration.
Although the two sets of migration tracks shown in Figure~\ref{fig8} are
very different, we do not see photoevaporation
significantly affecting the planets' migration
in either set.  As in the previous experiments presented in this paper,
photoevaporation has little effect on the planets' semimajor axes after
200,000~years of migration.

\begin{figure}
\epsscale{0.9}
\plotone{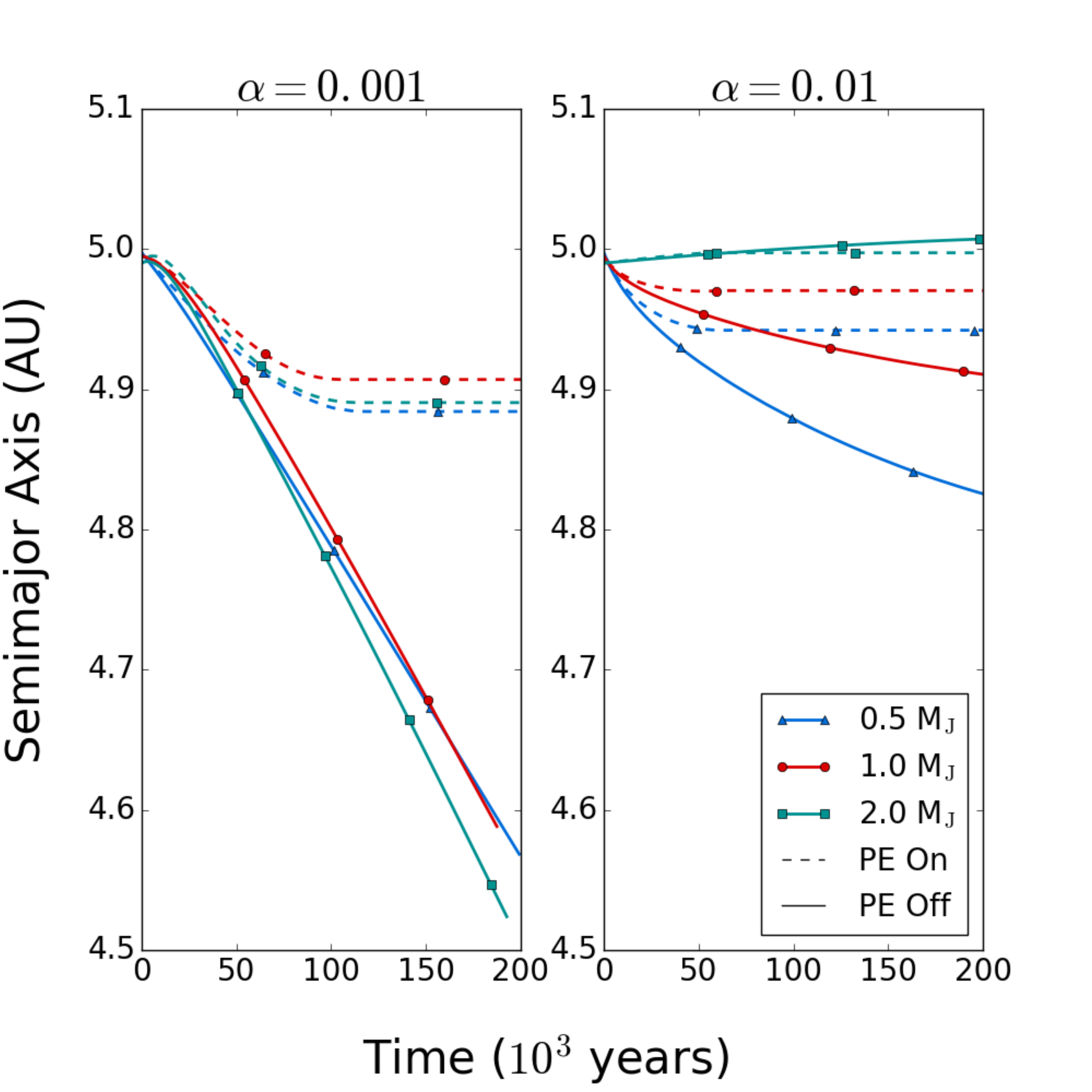}
\caption{Migration tracks for ER15 Comparison models with $M_\mathrm{disk}$=0.0007 M$_\sun$. The low disk mass results in the planets' inability to migrate significantly (note the change in y-axis scale from Figure~\ref{fig8}), and amplifies the relative effect of photoevaporation removing disk material. \label{fig9}}
\end{figure}

We further extend the ER15 Comparison parameter space to lower initial
disk mass in Figure~\ref{fig9}. Here we are simulating disks with
$t_{\rm pe} \la t_{\nu}$, so we see sharp surface density
gradients sculpted by photoevaporation. Initially, the planets in
photoevaporating disks migrate at nearly the same rates as the planets
in the control simulations. Then, between 50,000 and 100,000 years after
the start of migration, photoevaporation dissipates almost all of the
remaining disk gas. The planets hardly migrate at all after $t =
0.1$~Myr as there is very little mass in the disk to torque them, as
also found by \citet{l10} for planets with $M > 10 M_{\oplus}$. A
limitation of our ER15 Comparison/Extension results is that we do not
directly simulate the gradual depletion that would turn the disks from
Figure~\ref{fig8} ($M = 0.007 M_{\odot}$) into the disks from
Figure~\ref{fig9} ($M = 0.0007 M_{\odot}$), or the planet migration
during this depletion epoch. Due to computational time constraints, we
may be missing an epoch in between the time periods studied in
Figures~\ref{fig8} and \ref{fig9} during which photoevaporation might
gradually begin to slow giant planet migration. 
Still, these results suggest that the corotation torque, which acts
during the entire migration epoch ($>$100,000 yr - 2 Myr) for the
planet/disk parameters simulated here \citep[e.g.][]{cm07,dk15}, has more influence on migration tracks than photoevaporation. Since photoevaporation only sculpts steep surface density gradients during the final $\sim$100,000~years before disk dispersal, by which time disk masses are low enough to have nearly halted migration, we find that photoevaporation may only modify final planet locations by perhaps a few tenths of an AU, even for very late-forming planets as in AP12 and ER15.

\begin{figure}
\epsscale{1.0}
\plotone{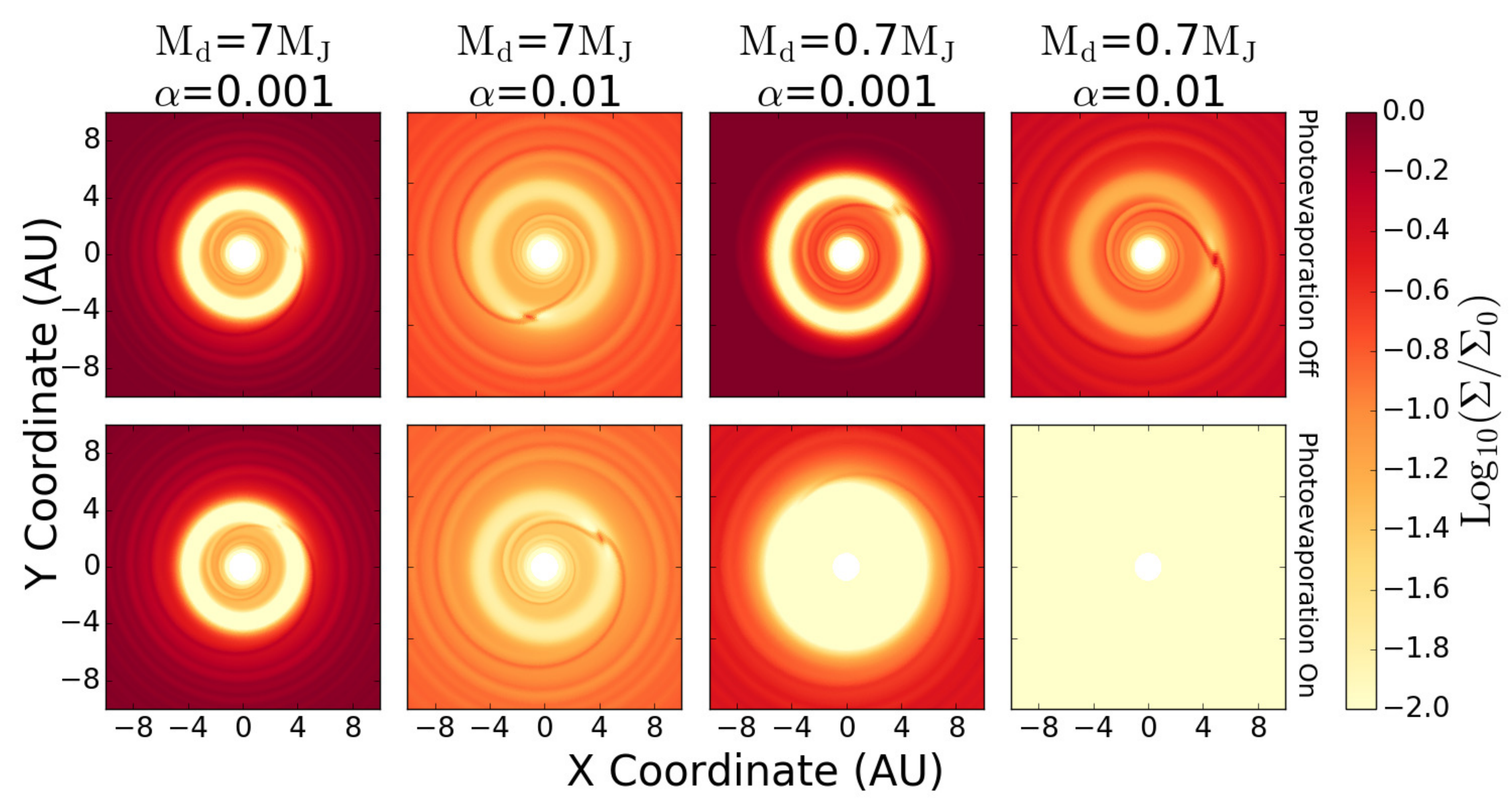}
\caption{Density profiles for ER15 Comparison simulations after $10^5$~yr. Low viscosity ($\alpha$=0.001) simulations show a significantly depleted inner disk due to the planet inhibiting accretion across its orbit, while high viscosity simulations show more overall disk depletion due to more accretion onto the central star. Disk structure only appears significantly affected by photoevaporation in the 0.7 M$_\mathrm{J}$ disks, where the disk interior to the planet is completely removed within 0.1~Myr.  \label{fig10}}
\end{figure}

Since ER15 did not map individual migration tracks, we cannot be sure
which part of their parameter space gave rise to the 1-2~AU planet pileup
found in their models. 
The pileup they predict is likely due to the PIPE mechanism, where giant planets inhibit accretion across their tidal gaps, starving the interior disk of material and speeding up photoevaporative clearing by opening up the outer disk to the direct field \citep{aa09, ap12, r13x}.
To see why, we examine Figure~\ref{fig10}, where we
plot the ER15 Comparison and Extension models' normalized surface
density distributions, $\Sigma(r,\theta,t) / \Sigma(r,\theta,0)$, at $t
= 0.1$~Myr (when the migration tracks in the left panel of
Figure~\ref{fig9} flatten).
In the disks with $M_0 = 0.007 M_{\odot}$
($7 M_{\rm J}$), photoevaporation has little accumulated effect on the
surface density distribution, even after 0.1~Myr of evolution.
For disks with $M_0 = 0.0007 M_{\odot}$ ($0.7 M_{\rm J}$),
photoevaporation's modifications to the surface density distributions
are obvious. In the weakly turbulent disk with $\alpha = 0.001$, almost
all material interior to the tidal gap has evaporated or accreted onto
the star without being replenished,
as photoevaporation and the planet's torque on the disk both inhibit material from accreting across the planet's tidal gap and replenishing the inner disk.
However, our simulations suggest that PIPE should have little effect on the planets' overall migration:
we find significant gas depletion well before the direct X-ray field would reach the disk exterior to the planet's orbit (though we do not actually model the direct field),
so the planets are hardly moving by the time the inner disk drains (note the small range of semimajor axes on the vertical axis of Figure~\ref{fig9}).

In the disk with $M = 0.7 M_J$ and $\alpha = 0.01$, viscosity moves
material from the outer to the inner disk more efficiently, helping
photoevaporation to drain the {\it entire} modeled region within 0.1~Myr
(Figure~\ref{fig10}). Here, too, gas densities are too low to drive
migration. Roughly speaking, the planet must interact with approximately
its own mass in disk gas in order to migrate significantly; the plots in
Figure~\ref{fig10} confirm the intuitive result that disks with less
mass than the planets they host cannot drive migration.

\section{Discussion and Conclusions} \label{discussion}

We have investigated the interaction between protoplanetary disk
photoevaporation and giant planet migration using 2-D disk+planet
models. In contrast to previous models claiming a
photoevaporation-induced pileup of giant planets near $\sim$1 AU
\citep{ap12,er15}, we directly compare simulations with photoevaporation
on and off to assess its impact. Any interaction between
photoevaporation and planet migration is
expected to be too small to be visible in the current
semimajor axis distribution of exoplanets based on our results:
\begin{enumerate}
\item{
When disk densities get too low for giant planets to form, viscous
forces continue to dominate
planet-disk interactions
over photoevaporation for $\sim$2~Myr.}
\item{During the vast majority of a giant planet's migration,
photoevaporation does not create any time- or location-specific
perturbations in its migration track.}
\item{The highest photoevaporation-rate model we tested (${\sim}$3
$\times 10^{-8}$~M$_\sun$~yr$^{-1}$; FUV) results in less than 5\% change in
final semimajor axis for a recently-formed Jupiter after 0.1~Myr
of migration, which is entirely due to the extra disk mass loss.}
\item{By the time photoevaporation can create steep enough gradients in the
disk surface density to significantly perturb migration rates,
migration is very slow, and may only continue for a few tenths of an AU before gas disk dissipation.}
\end{enumerate}

However, our conclusions come with the caveat that, due to limits on
computing time, we did not sample the space of significant parameters
(i.e. planet mass, planet formation location, disk viscosity, and disk
mass) as fully as we would like to make conclusions about the entire
population of giant exoplanets. While we were unable to find any cases
of disk photoevaporation significantly altering Type II migration rates
in our simulations, there may be pockets of parameter space with more
significant interactions that we missed. Small effects such as depletion
of co-rotating material in the tidal gaps of planets in low-viscosity
disks (\S~\ref{FOGSresults}) may be more significant in the unexplored
parameter space than we found in our simulations.
Future work with more computational resources could further clarify the
effects of photoevaporation on corotation torque.

Another caveat to our conclusions is that in 2-D, as opposed to 3-D,
the vertical disk density
profile is not modified by the presence of the giant planet or
photoevaporation.  Photoevaporation would almost certainly modify the
vertical disk structure near the planet since it expels material that is
highest above the midplane, and the most significant vertical density
sculpting caused by photoevaporation would occur in the depleted
co-orbital region due to its low surface density. Vertical density
sculpting may alter the local disk temperature profile by changing the
height above the midplane at which stellar radiation is absorbed
\citep[e.g.][]{j08,jt12,jt13}, a process which we have not explored. By
linking a photoevaporation model with a 3-D simulation of planet
migration such as those performed by \cite{f17} for small planets,
one could explore the effects of photoevaporation on
vertical disk structure near a planet, though such a simulation would be
computationally expensive. Also, since the photoevaporation models we
used assume a constant stellar UV/X-ray radiation field and were derived
for disks without planets, our disk mass loss profile is azimuthally
symmetric and does not vary over time.  However, disk density structures
such as a planet's tidal gap, horseshoe region and tidal tails may
significantly alter local photoevaporation rates.

Our 2-D simulations have the advantage over 1-D simulations by AP12 and
ER15 in that we model the effects of photoevaporation on non-axisymmetric
gas inside the planet's tidal gap, especially corotating material. Furthermore, our 2-D simulations of viscous disks naturally include gas accretion across the planet's tidal gap \citep{dk15}, forming tidal tails that allow stars that host planetary systems to still accrete gas \citep{ds11,d16}. Although the 1-D migration torque formula used by AP12 and ER15 treats the tidal gap walls as impermeable, AP12 and ER15 mimic a gas-permeable gap with an accreting planet by using coupled parameters that describe gap-crossing efficiency and planet accretion efficiency, and are functions of planet mass and disk viscosity \citep{va04}. In both studies, the efficiency parameters significantly affect the synthesized semimajor axis distributions. We do not consider planet accretion, which could reduce gas flow across the tidal gap \citep{d02,b03,wc10,dk17} and speed up viscous depletion of the inner disk. It is possible that the PIPE process \citep{r13x} halts migration for many gas giants, an effect we have not captured here.


AP12 and ER15 do not include comparison
simulations with photoevaporation off, so the exact effect of
photoevaporation on their semimajor axis distributions is not known. However,
AP12 convincingly argue that the deserts and pileups in their semimajor
axis distributions are caused by planetary tidal gaps suppressing
accretion to the inner disk, allowing it to drain quickly and triggering
the UV-switch where direct-field EUV photoevaporation \citep{a06a,a06b} quickly
removes the outer disk \citep[for a more detailed explanation of
this mechanism, called PIPE, see][]{aa09,r13x}. ER15 use the same UV-switch,
so features in their semimajor axis distributions are also probably due to the PIPE mechanism.
The semi-major axis distributions predicted by AP12 and ER15 appear
roughly consistent with our findings that a marginal effect of PIPE
may be overshadowed by a long epoch of migration.
The presence of a long migration epoch casts doubt on whether
photoevaporation can leave a signature on the semimajor axis distribution of
giant planets, as photoevaporation cannot affect giant
planet migration at late times if migration has already stalled due to
other processes. In high-viscosity disks, corotation torque can slow or
even reverse migration \citep{cm07,p08,pp09,d14,p14}. Tidal
circularization of giant planets can halt migration at very small radii
\citep{pb13}. 
Traps for giant planet cores
created by ice lines, dead zones, and heat transitions may determine
giant planet formation locations, leaving more significant signatures on
the final giant planet semimajor axis distribution
\citep{m06,kl07,s11,hp11,hp12,hp13,hp14}.
In multi-planet systems, orbital
migration from planet-planet scattering may dominate over migration due
to planet-disk interactions \citep{f01,fr08,bn12,ma12}.

Finally, we note that the observed pileup in the semimajor axis distribution
of exoplanets around 1~AU \citep{us07,w09,hp12,bn13} loses much of its
strong visual presence when the distribution is plotted on a linear
semimajor axis scale.
Considering the complex nature of planet migration and
its many theorized halting mechanisms, we question the usefulness of plotting
planet frequency on a
log semimajor axis scale when testing migration theory.
If photoevaporation does
have a significant effect on the semimajor axis distribution of
exoplanets, such an effect must exist outside of both the parameter
space of our simulations and the known properties of confirmed exoplanets.

\acknowledgments

The authors thank Frederic Masset for making FARGO publicly
available online. We thank Ralph Pudritz for inspiring this project via
a talk at IAU symposium 299, and Yasuhiro Hasegawa for insights that
helped greatly improve this work and place it in the context of the
planet migration field. We thank James Owen, David Tsang, and Eric Ford
for contributing ideas that helped shaped this work. We acknowledge
support from the University of Delaware Department of Physics and
Astronomy for providing computing resources through the Farber and Mills
computing clusters at University of Delaware. We also acknowledge
support from Dr.\ Dodson-Robinson's start-up grant from the UNIDEL
foundation, and from NSF CAREER award 1520101.

\bibliography{photoevaporation}
\clearpage





\appendix

\section{Modifications to FARGO} \label{fargomod}

\begin{figure}
\centering
\includegraphics[scale=0.4]{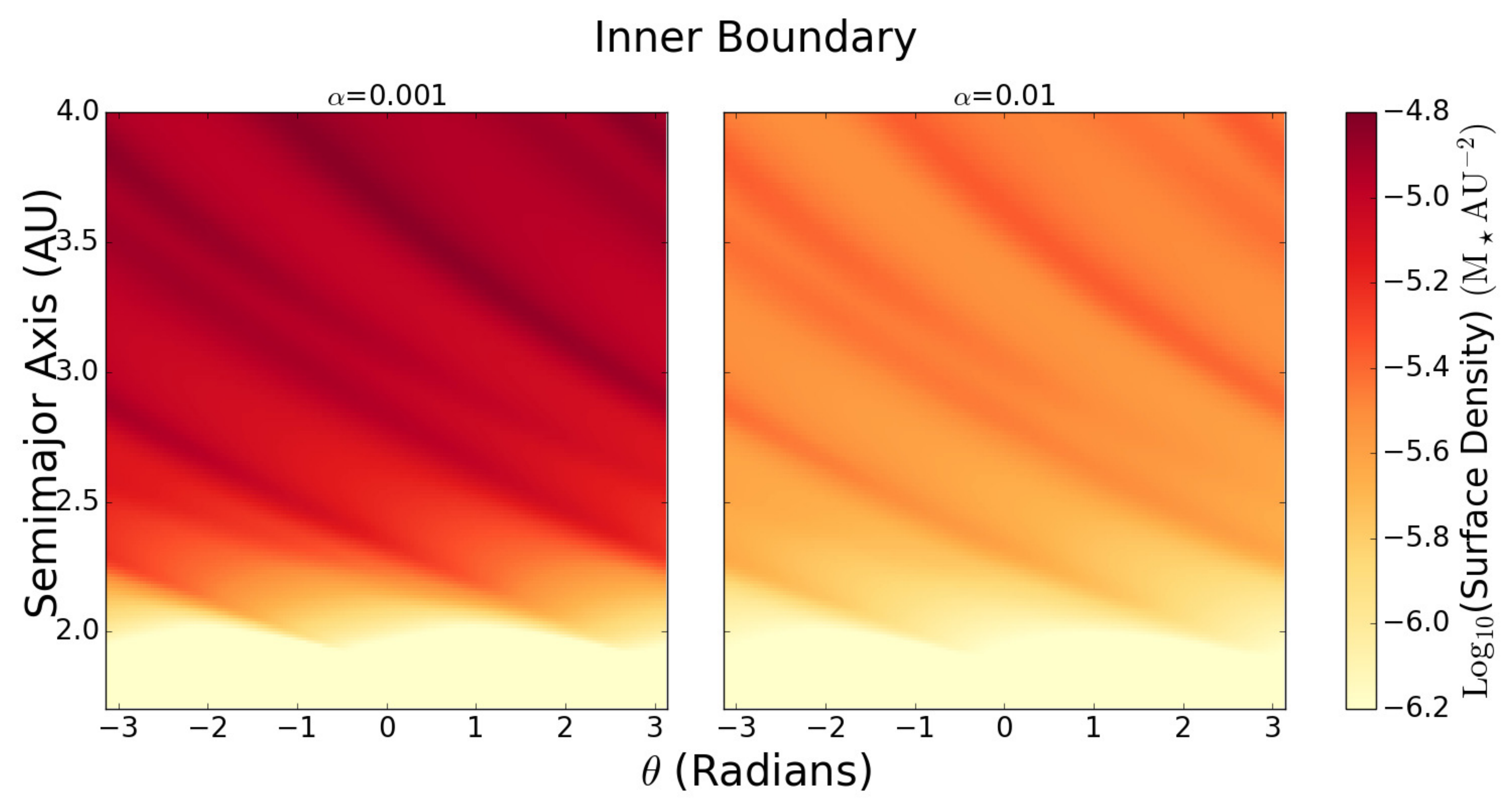}
\includegraphics[scale=0.4]{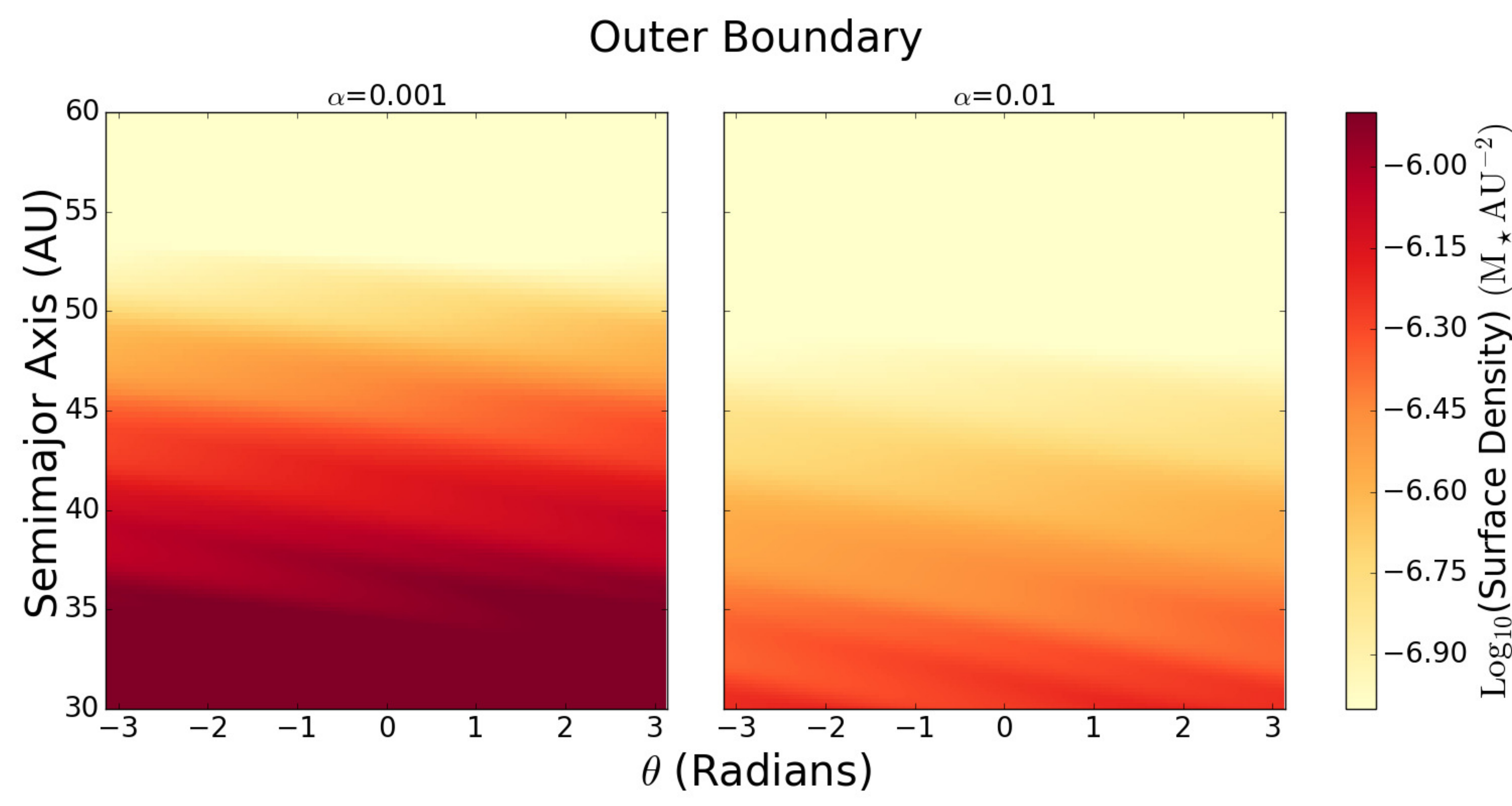}
\caption{Density waves created by the planet interacting with the inner (top) and outer (bottom) boundaries for $\alpha = 0.01$ and $\alpha = 0.001$ in our Fixed Orbit simulations after 50 kyr. Though no active wave damping has been imposed at the boundaries, our boundary conditions do not allow reflected waves to travel away from the boundaries.
\label{fig14}}
\end{figure}

FARGO 2-D is a polar mesh hydro code that uses finite differencing to
solve the Navier-Stokes equations for a Keplerian disk using a full
viscous stress tensor. It considers the gravity of the central object as
well as any number of planets, but no self-gravity of the disk. An
isothermal equation of state is used with an arbitrary radial
temperature profile.  Advection is accomplished using the van Leer
upwinding technique on a staggered mesh along with the FARGO (Fast
Advection in Rotating Gaseous Objects) algorithm. We added
``zero-torque" boundary conditions \citep{a06b} to the inner and outer
disk boundaries by setting the inner and outer zone surface densities to
$10^{-21} \mathrm{M_\sun\ AU^{-2}}$. To prevent negative densities from arising, we
set the radial velocity to 0 for any empty zone at the disk boundary. We
increase the von Neumann-Richtmyer viscosity constant, the number of
zones over which shocks are spread, to 3.41 instead of the FARGO default
of 1.41 in order to avoid crashes caused by discontinuities at the
boundaries.
Figure~\ref{fig14} shows how density waves produced by the
planet interact with the inner (top) and outer (bottom) boundaries.
If density waves were reflecting off the boundaries, we would expect
to see a cross-hatching pattern in Figure~\ref{fig14} similar to Figure 19
in \cite{dV06}. Instead, our zero-density boundary conditions swallow up potential
density waves, allowing them to flow out of the disk, but preventing reflected
waves from re-entering the disk as there is no material in the boundary
zones to act as a wave source. As we do not find reflected wave patterns
traveling back toward the planet's orbit \citep[e.g. as in ][]{dV06},
we do not need to impose active wave-damping at the boundaries.

Previous studies combining photoevaporation with 2-D FARGO simulations
\citep{ma12,r13x,r15} use an `open' boundary condition on the inner disk
boundary where the disk surface density is set to its initial value.
This open boundary condition limits viscous draining onto
the star since the ring of zones just outside the boundary ring can lose material,
but the boundary ring keeps being reset to the initial density value, resulting in
a density gradient at the boundary that prevents accretion onto the star
after 0.2 Myr or so. This difference in viscous disk draining makes it difficult
to directly compare their disks' dispersal with ours in detail, since our disks
are still accreting onto the central star (in addition to being photoevaporated) 
even at very low disk masses.

Published prescriptions for photoevaporation are described in
\S~\ref{photoevap}. To implement them, we use an azimuthally symmetric
array of $\dot{\Sigma}(r)$ values computed from each one of the
published profiles. For each FARGO time step $\Delta t$, the surface
density subtracted from each zone is $\dot{\Sigma}(r) \Delta t$. To
prevent negative density values and other numerical instabilities, we
skip the density subtraction that represents photoevaporation in any
zone where it would remove more than 10\% of the surface
density. In practice this condition is rarely met, as the
disk surface density in a given zone must be extremely low for the
density subtraction to be skipped. By the time our simulations
reach such low densities, the photoevaporating radiation would
be passing through an optically thin disk, rendering the
published expressions for $\dot{\Sigma}(r)$ invalid.

\section{Simulation Parameters Tables} \label{AppendixParams}

\begin{table}
\caption{Parameters for All Disk+Planet Simulations}
\tabletypesize{\tiny}
\centering
\begin{adjustbox}{width=1.1\textwidth, center=\textwidth}
\begin{tabular}{lllllllll}
\tableline\tableline
Disk Model & PE Model & $\mathrm{M_{disk}(t=0)}$ [$\mathrm{M_{sun}}$] & $\mathrm{M_{planet}}$ [$\mathrm{M_{Jupiter}}$] &  $\mathrm{a_{planet}(t=0)}$ [$\mathrm{AU}$] & Alpha &  $\mathrm{R_{in}}$ [$\mathrm{AU}$] & $\mathrm{R_{out}}$ [$\mathrm{AU}$] & H / dr\tablenotemark{a} \\
\tableline
Planet Forming Disk & EUV & 0.07209 & 0.05\tablenotemark{b}, 1.0 & 1.049 & 0.01 & 0.3405 & 204.3 & 3.8 \\
Planet Forming Disk & EUV & 0.07209 & 0.05\tablenotemark{b}, 1.0 & 1.573 & 0.01 & 0.3405 & 204.3 & 4.2 \\
Planet Forming Disk & EUV & 0.07209 & 0.05\tablenotemark{b}, 1.0 & 2.132 & 0.01 & 0.3405 & 204.3 & 4.6 \\
Planet Forming Disk & X-ray & 0.07161 & 1.0 & 6.640  & 0.001 & 1.703 & 204.3 & 8.1 \\
Planet Forming Disk & X-ray & 0.07161 & 1.0 & 9.943 & 0.001 & 1.703 & 204.3 & 8.9 \\
Planet Forming Disk & X-ray & 0.07161 & 1.0 & 13.28 & 0.001 & 1.703 & 204.3 & 9.6 \\
Planet Forming Disk & X-ray & 0.07161 & 0.05\tablenotemark{b}, 1.0 & 6.640  & 0.01 & 1.703 & 204.3 & 8.1 \\
Planet Forming Disk & X-ray & 0.07161 & 0.05\tablenotemark{b}, 1.0 & 9.943 & 0.01 & 1.703 & 204.3 & 8.9 \\
Planet Forming Disk & X-ray & 0.07161 & 0.05\tablenotemark{b}, 1.0 & 13.28 & 0.01 & 1.703 & 204.3 & 9.6 \\
Planet Forming Disk & FUV & 0.07161 & 0.05\tablenotemark{b}, 1.0 & 7.900 & 0.01 & 1.703 & 204.3 & 8.4 \\
Planet Forming Disk & FUV & 0.07161 & 0.05\tablenotemark{b}, 1.0 & 11.85 & 0.01 & 1.703 & 204.3 & 9.3 \\
Planet Forming Disk & FUV & 0.07161 & 0.05\tablenotemark{b}, 1.0 & 15.80 & 0.01 & 1.703 & 204.3 & 10.0 \\
\tableline
Fixed Orbit & EUV & 0.003490 & 1.0 & 1.049 & 0.001, 0.01 & 0.3405 & 10.22 & 7.2 \\
Fixed Orbit & EUV & 0.003490 & 1.0 & 1.573 & 0.001, 0.01 & 0.3405 & 10.22 & 7.9 \\
Fixed Orbit & EUV & 0.003490 & 1.0 & 2.097 & 0.001, 0.01 & 0.3405 & 10.22 & 8.5 \\
Fixed Orbit & X-ray & 0.02347 & 1.0 & 6.640 & 0.001, 0.01 & 1.703 & 68.10 & 10.5 \\
Fixed Orbit & X-ray & 0.02347 & 1.0 & 9.943 & 0.001, 0.01 & 1.703 & 68.10 & 11.6 \\
Fixed Orbit & X-ray & 0.02347 & 1.0 & 13.28 & 0.001, 0.01 & 1.703 & 68.10 & 12.5 \\
Fixed Orbit & FUV & 0.02347 & 1.0 & 7.900 & 0.001, 0.01 & 1.703 & 68.10 & 11.0 \\
Fixed Orbit & FUV & 0.02347 & 1.0 & 11.85 & 0.001, 0.01 & 1.703 & 68.10 & 12.1 \\
Fixed Orbit & FUV & 0.02347 & 1.0 & 15.80 & 0.001, 0.01 & 1.703 & 68.10 & 13.0 \\
\tableline
ER15 Comparison & X-ray & 0.007 & 0.5 & 5.0 & 0.0009545 & 1.022 & 68.10 & 8.6 \\
ER15 Comparison & X-ray & 0.007 & 1.0 & 5.0 & 0.0009545 & 1.022 & 68.10 & 8.6 \\
ER15 Comparison & X-ray & 0.007 & 2.0 & 5.0 & 0.0009545 & 1.022 & 68.10 & 8.6 \\
ER15 Extension & X-ray & 0.007 & 0.5 & 5.0 & 0.009545 & 1.022 & 68.10 & 8.6 \\
ER15 Extension & X-ray & 0.007 & 1.0 & 5.0 & 0.009545 & 1.022 & 68.10 & 8.6 \\
ER15 Extension & X-ray & 0.007 & 2.0 & 5.0 & 0.009545 & 1.022 & 68.10 & 8.6 \\
ER15 Extension & X-ray & 0.0007 & 0.5 & 5.0 & 0.0009545 & 1.022 & 68.10 & 8.6 \\
ER15 Extension & X-ray & 0.0007 & 1.0 & 5.0 & 0.0009545 & 1.022 & 68.10 & 8.6 \\
ER15 Extension & X-ray & 0.0007 & 2.0 & 5.0 & 0.0009545 & 1.022 & 68.10 & 8.6 \\
ER15 Extension & X-ray & 0.0007 & 0.5 & 5.0 & 0.009545 & 1.022 & 68.10 & 8.6 \\
ER15 Extension & X-ray & 0.0007 & 1.0 & 5.0 & 0.009545 & 1.022 & 68.10 & 8.6 \\
ER15 Extension & X-ray & 0.0007 & 2.0 & 5.0 & 0.009545 & 1.022 & 68.10 & 8.6 \\
\tableline
Crida \& Morbidelli Extension & None & 0.02371 & 1.0 & 5.0 & 0.005 & 1.022 & 68.10 & 8.6 \\
Crida \& Morbidelli Extension & None & 0.02371 & 1.0 & 5.0 & 0.01 & 1.022 & 68.10 & 8.6 \\
Crida \& Morbidelli Extension & None & 0.02371 & 1.0 & 5.0 & 0.02 & 1.022 & 68.10 & 8.6 \\
Crida \& Morbidelli Extension & None & 0.02371 & 1.0 & 5.0 & 0.05 & 1.022 & 68.10 & 8.6 \\
Crida \& Morbidelli Extension & None & 0.02371 & 1.0 & 10.0 & 0.005 & 1.022 & 68.10 & 10.2 \\
Crida \& Morbidelli Extension & None & 0.02371 & 1.0 & 10.0 & 0.01 & 1.022 & 68.10 & 10.2 \\
Crida \& Morbidelli Extension & None & 0.02371 & 1.0 & 10.0 & 0.02 & 1.022 & 68.10 & 10.2 \\
Crida \& Morbidelli Extension & None & 0.02371 & 1.0 & 10.0 & 0.05 & 1.022 & 68.10 & 10.2 \\
\tableline
\end{tabular}
\end{adjustbox}
\tablenotetext{a}{Effective resolution: Disk scale height (H) divided by radial zone width (dr), evaluated at t=0 planet orbital radius.}
\tablenotetext{b}{Neptune-mass planets were simulated in disks with an aspect ratio of 0.0386 instead of 0.0546, an inner boundary of 0.3405~AU, and 400 radial grid zones.}
\tablecomments{See Table~\ref{PFDparams} for an example FARGO parameters file. For the boundary conditions we used, see Appendix~\ref{fargomod}.} \label{params}
\end{table}

\clearpage

\begin{table}
\centering
\caption{Example FARGO Parameters File\label{PFDparams}}
\begin{tabular}{ll}
\tableline\tableline
FARGO Parameter & Value \\
\tableline
AspectRatio &0.05455610099\\
Sigma0 &1.91549508e-4\\
AlphaViscosity &0.01\\
SigmaSlope &1.0\\
FlaringIndex &0.25\\
ThicknessSmoothing &0.6\\
InnerBoundary &Custom\tablenotemark{a}\\
OuterSourceMass &Custom\tablenotemark{a}\\
Frame &Fixed\\
Nrad &600\\
Nsec &200\\
Rmin &0.5\tablenotemark{b}\\
Rmax &60.0\tablenotemark{b}\\
RadialSpacing &Logarithmic\\
\tableline
\end{tabular}
\tablenotetext{a}{See Appendix~\ref{fargomod} for the boundary conditions we implemented}
\tablenotetext{b}{Note the conversion between FARGO units and AU is $(2 \pi )^{-2/3}$}
\tablecomments{If a parameter is not specified, the default value was used. See the online documentation at http://fargo.in2p3.fr/-Parameters- for parameter descriptions.}
\end{table}


\section{Additional Migration Tracks} \label{AppendixMigration}

\begin{figure}[h]
\epsscale{0.65}
\plotone{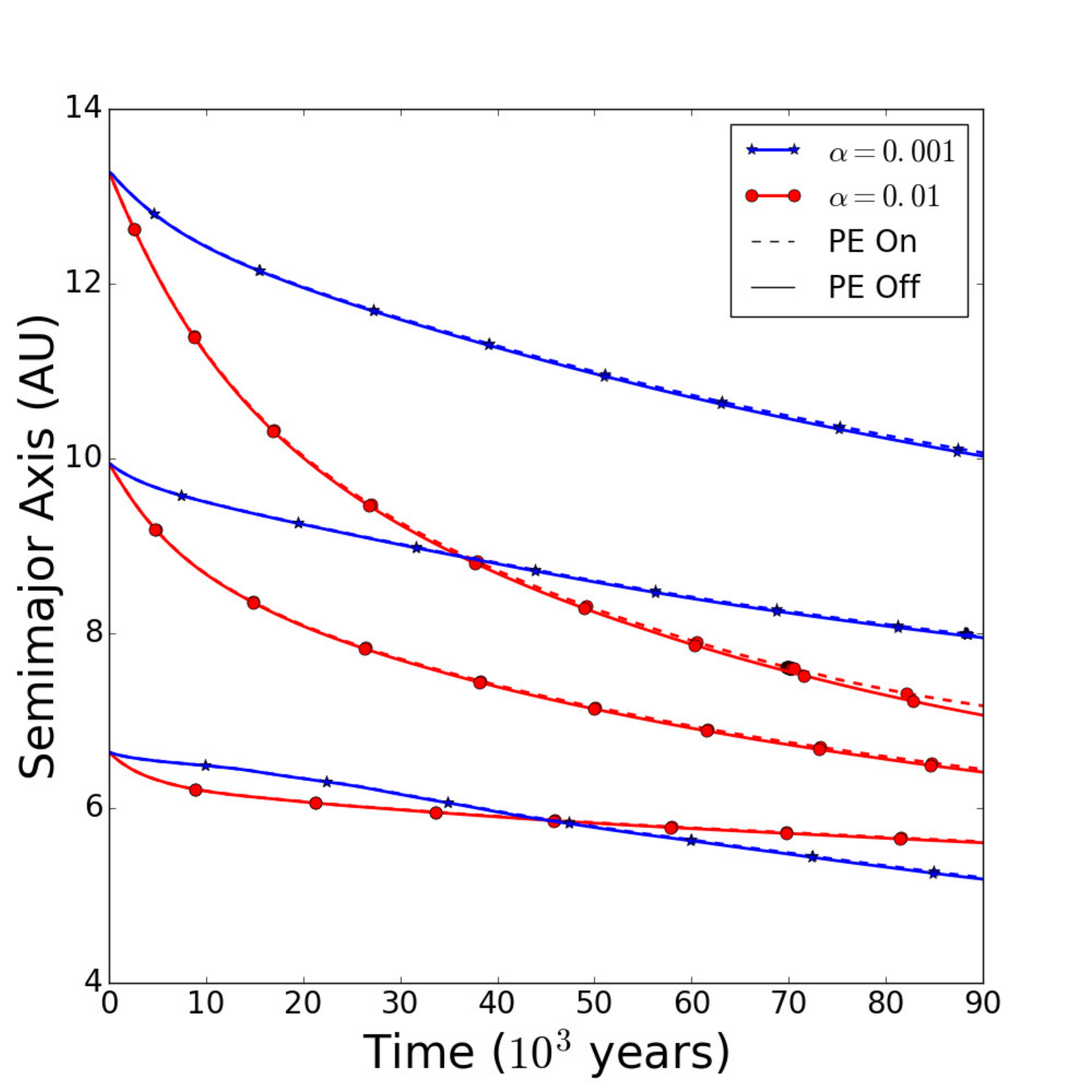}
\caption{Migration tracks of Jupiter-mass planets with starting locations centered around where X-ray photoevaporation (PE) opens its gap, for $\alpha=0.01$ and $\alpha=0.001$ disks. The red curves are identical to those in Figure~\ref{fig5}; we include this figure to highlight the dominant effect of disk viscosity on planet migration. \label{fig15}}
\end{figure}








\end{document}